\documentclass{article}

\usepackage{epsfig} 
\usepackage{amsmath}
\usepackage{amsfonts}
\usepackage{graphicx}

\date{\today}

\newcommand{\insertplot}[5]{\begin{figure}
 \hfill\hbox to 0.05in{\vbox to #5in{\vfill
 \inputplot{#1}{#4}{#5}}\hfill}
 \hfill\vspace{-.1in}
 \caption{#2}\label{#3}
 \end{figure}}
 \newcommand{\inputplot}[3]{
 \special{ps: plotfile #1}
}
\newcounter{fig}

\begin{document}

\title{AdS$_5$ solutions in 
Einstein--Yang-Mills--Chern-Simons theory} 
  
\author{{\large Yves Brihaye,}$^{\dagger}$
{\large Eugen Radu}$^{\ddagger}$
and {\large D. H. Tchrakian}$^{\star \diamond }$ \\  
$^{\dagger}${\small Physique-Math\'ematique, Universit\'e de Mons, Mons, Belgium}\\
$^{\ddagger}${\small  Institut f\"ur Physik, Universit\"at Oldenburg, Postfach 2503
D-26111 Oldenburg, Germany}
  \\
$^{\star}${\small  Department of Computer Science,
National University of Ireland Maynooth,
Maynooth,
Ireland} \\
$^{\diamond}${\small School of Theoretical Physics -- DIAS, 10 Burlington
Road, Dublin 4, Ireland }}

\newcommand{\ta}{\theta}
\newcommand{\Si}{\Sigma}
\newcommand{\vf}{\varphi}
\newcommand{\dd}{\mbox{d}}
\newcommand{\tr}{\mbox{tr}}
\newcommand{\la}{\lambda}
\newcommand{\ka}{\kappa}
\newcommand{\f}{\phi}
\newcommand{\al}{\alpha}
\newcommand{\ga}{\gamma}
\newcommand{\de}{\delta}
\newcommand{\si}{\sigma}
\newcommand{\bomega}{\mbox{\boldmath $\omega$}}
\newcommand{\bsi}{\mbox{\boldmath $\sigma$}}
\newcommand{\bchi}{\mbox{\boldmath $\chi$}}
\newcommand{\bal}{\mbox{\boldmath $\alpha$}}
\newcommand{\bpsi}{\mbox{\boldmath $\psi$}}
\newcommand{\brho}{\mbox{\boldmath $\varrho$}}
\newcommand{\beps}{\mbox{\boldmath $\varepsilon$}}
\newcommand{\bxi}{\mbox{\boldmath $\xi$}}
\newcommand{\bbeta}{\mbox{\boldmath $\beta$}}
\newcommand{\ee}{\end{equation}}
\newcommand{\eea}{\end{eqnarray}}
\newcommand{\be}{\begin{equation}}
\newcommand{\bea}{\begin{eqnarray}}

\newcommand{\ii}{\mbox{i}}
\newcommand{\e}{\mbox{e}}
\newcommand{\pa}{\partial}
\newcommand{\Om}{\Omega}
\newcommand{\vep}{\varepsilon}
\newcommand{\bfph}{{\bf \phi}}
\newcommand{\lm}{\lambda}
\def\theequation{\arabic{equation}}
\renewcommand{\thefootnote}{\fnsymbol{footnote}}
\newcommand{\re}[1]{(\ref{#1})}
\newcommand{\R}{{\rm I \hspace{-0.52ex} R}}
\newcommand{\N}{{\sf N\hspace*{-1.0ex}\rule{0.15ex}%
{1.3ex}\hspace*{1.0ex}}}
\newcommand{\Q}{{\sf Q\hspace*{-1.1ex}\rule{0.15ex}%
{1.5ex}\hspace*{1.1ex}}}
\newcommand{\C}{{\sf C\hspace*{-0.9ex}\rule{0.15ex}%
{1.3ex}\hspace*{0.9ex}}}
\newcommand{\eins}{1\hspace{-0.56ex}{\rm I}}
\renewcommand{\thefootnote}{\arabic{footnote}}

\maketitle

\begin{abstract}
We investigate static, spherically symmetric solutions 
of an Einstein-Yang-Mills-Chern-Simons system with negative cosmological constant,
for an $SO(6)$ gauge group. 
For a particular value of the  Chern-Simons coefficient, this model can be viewed
as a truncation of the five-dimensional maximal gauged supergravity and we expect that the basic
properties  of the solutions in the full model to persist in this truncation. 
Both globally regular, particle-like solutions and black holes are considered.
In contrast with the Abelian case, the  contribution of the  Chern-Simons term is nontrivial already
in the static,   spherically symmetric  limit. We find two types of solutions: the $generic$ configurations whose
magnetic gauge field
does not vanish fast enough at infinity (although the spacetime is asymptotically AdS), whose mass function is
divergent, and the $special$ configurations, whose existence depends on the Chern--Simons term, which are endowed with
finite mass. In the case of the $generic$ configurations, we argue that the divergent mass implies a nonvanishing trace
for the stress tensor of the dual $d=4$ theory.
 \end{abstract}
\medskip 
\medskip  
\section{Introduction}
 
  It was originally found in $d=4$ spacetime dimensions
\cite{Winstanley:1998sn}, \cite{Bjoraker:2000qd},  that 
a variety of well known features of asymptotically flat self-gravitating non-Abelian
solutions are not shared by their anti-de Sitter (AdS) counterparts.
In the presence of a negative cosmological constant $\Lambda<0$, the
Einstein-Yang-Mills (EYM) theory 
possesses a continuous spectrum of regular and black hole non-Abelian   
solutions in terms of the adjustable parameters that specify the initial conditions at the origin or at the event
horizon, rather than at  discrete values of these parameters.
The gauge field of generic solutions does not vanish asymptotically,
resulting in a nonzero magnetic flux at infinity.
Moreover, in contrast with the $\Lambda=0$ case,
some of the AdS configurations are stable against linear perturbations 
\cite{Breitenlohner:2003qj}.

As found in \cite{Okuyama:2002mh}, \cite{Radu:2005mj}, some of these features are shared 
by higher dimensional EYM solutions with AdS asymptotics.
Since gauged supergravity theories generically contain non-Abelian matter fields in
the bulk, these configurations are relevant in an AdS/CFT context, offering the possibility
of studying some aspects of the nonperturbative structure
of a CFT in a background gauge field  \cite{Mann:2006jc}.
On the CFT side, the boundary non-Abelian fields correspond  to external source currents coupled to various operators.
 
Given its relevance in the conjectured AdS/CFT correspondence \cite{Maldacena:1997re}, \cite{Witten:1998qj}, 
the case of ${\cal N}=8,~d=5$ gauged supergravity \cite{Gunaydin:1985cu}, \cite{Cvetic:2000nc} is of particular interest.
The bosonic sector of this theory consists of the metric, twenty scalars and fifteen $SO(6)$
Yang-Mills gauge fields\footnote{Note that the field content of the full ${\cal N}=8,~d=5$ gauged supergravity
is richer. However, a number of bosonic fields 
can be consistently  set to zero \cite{Cvetic:2000nc}.}.
Apart from the usual $F^2$ term, the Yang-Mills  (YM) fields have in this case a  non-Abelian 
Chern-Simons (CS) term  in the action, which  unlike in the Abelian case does not vanish when subjected
to spherical symmetry. 

Solutions of this model have been considered by several authors for various consistent truncations,
with subgroups of $SO(6)$ (see e.g. the recent work \cite{Cvetic:2009id} and the references therein).
However, to our knowledge, no attempt has been made to construct non-Abelian solutions
for the general case of the full $SO(6)$ gauge group.
In particular, the effects resulting from the introduction of
the CS term have so far not been studied.

This paper is aimed as a first step in this direction,
by taking a truncation of the  ${\cal N}=8,~d=5$  model
corresponding to a pure Einstein-Yang-Mills-Chern-Simons (EYMCS) theory
($i.e.$ with a negative cosmological constant, but with no  scalar fields).
We propose an Ansatz for a spherically symmetric $SO(6)$ gauge group 
and investigate the basic properties of both the black hole, and, particle-like globally regular solutions. 
Special attention is paid to the new features induced by the CS  term.  

As originally found in \cite{Volkov:2001tb}, 
\cite{Okuyama:2002mh}, \cite{Radu:2005mj}, a generic property of
higher dimensional EYM solutions is that their masses and actions,
as defined in the usual way, diverge. (For a recent review of these solutions, see \cite{Radu:2009rs}.) 
This can be understood heuristically by noting that the Derrick scaling requirement is not fulfilled in
spacetimes for dimension five and higher. 
To our knowledge, the only mechanism for regularising the mass of the $d>4$ asymptotically
flat or (A)dS non-Abelian gravitating
solutions, proposed so far in the literature, is to include higher order terms in the YM hierarchy
\cite{Brihaye:2002hr}, \cite{Radu:2005mj}, \cite{Brihaye:2006xc}
 and the corresponding YM--Higgs terms~\cite{Breitenlohner:2009zi}\footnote{ It is in principle possible to supply such higher scaling terms by employing only Higgs kinetic
terms, or, the kinetic terms of suitably gauged higher dimensional sigma models~\cite{LMP},
but these have not been attempted.}.
These are the 
YM counterparts of the Lovelock gravities (or the hierarchy of Einstein systems),
and occur in the low energy effective action of string theory~\cite{GSW,Pol}.

One of the main features of the
present work is the introduction of the CS term in $4+1$ dimensions, as an $alternative$ to the 
higher order curvature terms of the YM hierarchy employed previously for regularising the mass. It turns out that
this prescription $does$ result in finite mass solutions, but in addition to these, we find solutions with divergent
mass in what we have termed as the $generic\ \ case$. These last are characterised by a continuum of values of the
shooting parameters, which is a typical feature of EYM system with
a negative cosmological constant.  The finite mass solutions on the other hand,
termed as the $special\ \ case$, are special in that they occur only for a discrete set of values
of the shooting parameters. Of course, in the absence of the CS term the $only$ solutions that exist are ones with
divergent mass. 

Although the spacetime still approaches asymptotically the maximally symmetric AdS background,
the mass and the total action of a $generic$ solution present a logarithmically divergent part.
The  coefficient of the divergent term is fixed by the square of the induced non-Abelian fields on the boundary at
infinity. 
We shall argue that the logarithmic divergence of the non-Abelian AdS$_5$ configurations does not signal
a problem with these solutions, but rather provides a consistency check of the AdS/CFT conjecture, the coefficient
of the divergent term in the action being related in this case to the trace anomaly of the dual CFT
defined in a background non-Abelian field. Moreover, one can define a mass and action for the generic 
solutions by using the counterterm prescription of \cite{Balasubramanian:1999re}. 
The counterterms here depend not only on the boundary metric but also on the induced 
non-Abelian fields on the boundary.

However, perhaps the most interesing feature of the EYMCS model is the existence of a set of solutions with finite mass.
In the case of these solutions, as for the well known 
$d=4$ Bartnick-McKinnon solitons \cite{Bartnik:1988am}, they exist only for discrete values
of the shooting parameters (associated with the initial values of the gauge fields). 
As can be seen by using a simple Derrick-type argument, they are supported by the contribution of the non-Abelian CS term,
a prescription which can be exploited only in odd dimensional spacetimes where a CS term is defined.

The paper is structured as follows: in Section 2 we present the general framework and analyse the
field equations and boundary conditions. We present the numerical results in Section 3, special atention being paid to
 solutions with a finite mass. 
The  computation of the mass and electric charge of the solutions is addressed in Section 4.
We conclude with Section 5 where the   significance of, and further consequences arising from, the solutions we
have constructed are briefly discussed.

\section{The model}
\subsection{The action }

We consider the following action 
\bea
\label{action}
S= \int_{ \mathcal{M}}  d^5x \sqrt{-g} \left( \frac{1}{16\pi G} (R-2 \Lambda)-{\cal L}_{\rm{YM}} \right)
-\int_{ \mathcal{M}}  d^5x~{\cal L}_{\rm{CS}} ,
\eea
where 
\begin{eqnarray}
&&{\cal L}_{\rm{YM}}=
\frac14\,\mbox{Tr}\{ F_{\mu\nu}F^{\mu\nu} \},
\end{eqnarray}
is the usual Yang-Mills lagrangian for a gauge group $SO(6)$ 
(with $F_{\mu\nu}=\partial_\mu A_\nu-\partial_\nu A_\mu+e[A_\mu,A_\nu]$ the gauge field
strength tensor),
and 
\begin{eqnarray}
&&{\cal L}_{\rm{CS}} =i~ \kappa \,
\vep_{\mu\nu\rho\si\tau}\mbox{Tr}\, \bigg \{
A^{\tau}\left[F^{\mu\nu}F^{\rho\si}-
e F^{\mu\nu}A^{\rho}A^{\si}+
 \frac25 e^2 A^{\mu}A^{\nu}A^{\rho}A^{\si}\right]\, \bigg \},
\label{CS5}
\end{eqnarray}
is the CS term\footnote{The factor of $i$ appears in (\ref{CS5}) because we are using
an antihermitian representation for the $SO(6)$ algebra matrices.} 
(with $\kappa$ the CS coupling constant), while $\Lambda=-6/\ell^2$ is the cosmological constant
and $e$ is the gauge coupling constant.

These are basic the pieces which enters the bosonic action of the $d=5, ~{\cal N}=8$ 
gauged supergravity \cite{Gunaydin:1985cu}, \cite{Cvetic:2000nc}, the CS coefficient being $\kappa=1/8$ in this case.
However, the full ${\cal N}=8$ system contains in addition twenty scalars, which are represented
by a symmetric unimodular tensor. These scalars have a nontrivial potential approaching a constant negative value at
infinity which fixes the value of the effective cosmological constant.
Although ignoring the scalar sector is not a consistent trucation of the general ${\cal N}=8$ model,  
we expect that the basic
properties of our solutions hold also in that case\footnote{This is the situation for $d=4$ EYM solutions. 
As discussed $e.g.$ in \cite{Radu:2004xp}, the properties of the
EYM-dilaton solutions (the dilaton field possessing a nontrivial potential approaching a constant negative value at
infinity) are quite similar to those of the pure EYM-AdS case.}. 

The field equations are obtained by varying the action (\ref{action}) with
respect to the field variables $g_{\mu \nu},A_{\mu}$ 
\begin{eqnarray}
\label{Einstein-eqs}
R_{\mu \nu}-\frac{1}{2}g_{\mu \nu}R +\Lambda g_{\mu \nu}&=&
 8 \pi G ~T_{\mu \nu},
\\
\nonumber
\frac{1}{\sqrt{ -g}}D_{\mu}\left(\sqrt { -g}\,F^{\mu\tau}\right)+
3 \kappa \vep^{\mu\nu\rho\si\tau}\,F_{\mu\nu}\,F_{\rho\si}&=&0\,.
\end{eqnarray}
where the energy momentum tensor is defined by
\begin{eqnarray}
\label{Tij}
T_{\mu\nu} =
    \mbox{Tr}\bigg \{F_{\mu\alpha} F_{\nu\beta} g^{\alpha\beta}
   -\frac{1}{4} g_{\mu\nu}  ~ F_{\alpha\beta} F^{\alpha\beta}
  \bigg \}.
\end{eqnarray}
 One can show that this tensor is covariantly conserved ($i.e.$ $\nabla_\mu T^{\mu\nu}=0$)
 for solutions of the YMCS equations.

\subsection{The spherically symmetric Ansatz}
In this work we shall restrict to simplest case of static, spherically symmetric
solutions.
Thus we consider a  metric Ansatz in terms of two metric functions $N(r)$ and $\sigma(r)$
\begin{eqnarray}
\label{metric-gen} 
ds^{2}= 
 \frac{dr^2}{N(r)}+ r^2d\Omega^2_{3}-  N(r)\sigma^2(r) dt^{2},
\end{eqnarray}
where we have found convenient to define
\begin{eqnarray}
\label{formN}
N(r) =1-\frac {m(r)}{r^2}+\frac{r^2}{\ell^2}~,
\end{eqnarray}
the function $m(r)$ being related to the local mass-energy density (as defined in the standard way) up to some
 factor.
 $r$ and $t$ are the radial and time coordinates, while $d\Omega^2_{3}$ is the metric on the round three-sphere.

The static, spherically symmetric $SO(6)$ YM fields 
are taken in one  of the two chiral representations of $SO(6)$,
 such that the spherically
symmetric Ansatz is expressed in terms of the representation matrices,
\be
\label{sigma1}
\Sigma_{\al\beta}=-\frac14\Sigma_{[\al}\,\tilde\Sigma_{\beta]}\,,
\ee
 where 
 $\Sigma_{i}=-\tilde\Sigma_{i}=i\gamma_i\ ,\
\Sigma_{5}=-\tilde\Sigma_{5}=i\gamma_5\ ,\
\Sigma_{6}=+\tilde\Sigma_{6}=\eins$, 
are defined in terms of the usual Dirac gamma
matrices $\gamma_i$. 
 
Our spherically symmetric Ansatz for the $SO(6)$ YM connection $A_{\mu}=(A_t,A_i)$ is 
a variant of Witten's Ansatz for the axially symmetric instanton~\cite{Witten:1976ck}. Also, this Ansatz is one of
the two $SU(4)$ Ans\"atze proposed in \cite{Brihaye:1977qd}, namely the one employing Dirac gamma matrices as opposed
the one employing the Gell-Mann matrices\footnote{This distinction is important since the Dirac gamma matrix Ansatz
cannot be contracted to a $SU(2)$ subalgebra, while clearly  Gell-Mann matrix Ansatz does have a $SU(2)$ subalgebra.}. 
It is expressed as
\bea
A_t&=&\frac{1}{e}
\big( 
-\big(\vep\chi(r)\big)^M\,\hat x_j\,\Sigma_{jM}-
\chi^{7}(r)\,\Sigma_{56}
\big )
~,
\label{a02}
\\
\nonumber
A_i&=&\frac{1}{e}
\bigg( 
\left(\frac{\f^{7}(r)+1}{r}\right)\Sigma_{ij}\hat x_j+
\left[\left(\frac{\f^M(r)}{r}\right)\left(\delta_{ij}-\hat x_i\hat x_j\right)+
\big(\vep A_r(r)\big)^M\,\hat x_i\hat x_j\right]\Sigma_{jM}+
A_r^{7}(r)\,\hat x_i\,\Sigma_{56}
\bigg)~,
\label{ai2}
\eea
where $\hat x_i=x_i/r$  (with $x^i$ the usual Cartesian coordinates on $R^4$ and $x_ix_i=r^2$).
In the above relations $i,j=1,2,3,4$ and the index $M$ runs over $5,6$. Also, $\vep$ is the two
dimensional Levi-Civita symbol.

After taking the traces over the spin matrices, it is convenient to relabel the triplets of radial function as
$\vec\f\equiv(\f^M,\f^{3})$, $\vec\chi\equiv(\chi^M,\chi^{3})$ and $\vec A_r\equiv(A_r^M,A_r^{3})$, with $M=1,2$ now.

Substituting \re{a02}, \re{ai2} in the YM Lagrangian density
we find the compact expression for the reduced one dimensional reduced YM action density\footnote{Here we ignore
the (irrelevant) angular part in  $\sqrt{-g}$.}
\be
\label{YM}
{\cal L}_{\rm{YM}}\sqrt{-g}=
\frac{1}{e^2}
\bigg[
 \frac32\,r\,\si\left(N\,|D_r\f^a|^2+
\frac{1}{r^2}\left(|\f^a|^2-1\right)^2\right)
-\frac12\,\frac{r^3}{\si}\left(|D_r\chi^a|^2
+\frac{3}{Nr^2}\left(\vep^{abc}\f^b\chi^c\right)^2
\right)
\bigg]\, .
\ee
The calculation of the reduced CS action density is rather more tedious since
unlike \re{YM}, this term is not gauge invariant. It can be expressed in a compact way as
\bea
{\cal L}_{\rm{CS}} = &
  \kappa \frac{3}{4e^3}& 
\bigg[12(|\phi^d|^2-1)A^a\left(\vep^{abc}\chi^b\phi^c\right)
+ 3|\phi^b|^2\left( \f^a D_r\chi^a\right)    
\nonumber 
\\
&+& 6\,(\phi^b\chi^b)\,(\f^a D_r\f^a) 
-9|\f^b|^2\,(\chi^aD_r\f^a) 
\label{CS}  
  \\
&-&(2 \phi^3+5) (\f^aD_r\chi^a) 
+ (3|\phi^a|^2 - 2 \phi^3 - 5)\,D_r\chi^3 
\nonumber 
\\
&+&( 6\chi^3\phi^a+7\,\chi^a-2\phi^3\chi^a)\,D_r\f^a
-2(\chi^a\phi^a+\chi^3)\,D_r\phi^3\bigg]
 \,,
 \nonumber
\eea
which of course does not feature the metric functions.   Note that \re{CS} is not a scalar after contraction of the
indices $(a,b,c)$. This is a consequence of the $gauge \ variance$ of the CS density. 

In both \re{YM} and \re{CS} we have used the notation 
\be
\label{covp}
D_r\f^a=\pa_r\f^a+\vep^{abc}\,A_r^b\,\f^c\quad,\quad
D_r\chi^a=\pa_r\chi^a+\vep^{abc}\,A_r^b\,\chi^c~,
\ee
which are $SO(3)$ covariant derivatives of the two triplets
$\vec\f\equiv\f^a=(\f^M,\f^{3})$, and $\vec\chi\equiv\chi^a=(\chi^M,\chi^{3})$, with respect to the
$SO(3)$ gauge connection $\vec A_r\equiv A_r^a$. But $\vec A_r$ is really a $pure-gauge$ since in one dimension there
is no curvature. As such, it can be consistently set equal to zero. But more importantly, taking the variations
$\delta\vec A_r$ leads to the constraint equations, which are first
integrals of the equations for $\vec \phi$ and $\vec \chi$, and which play an important technical role in the numerical
integrations. We will return to these below. The ocurrence of constraint equations in a system supporting what are
basically $sphaleron$ solutions is completely expected, as the solutions we construct are indeed $sphalerons$,
just like the familiar Bartnik-McKinnon solutions. Needless to say, the consistency of the Ansatz used has been
verified, so it is sufficient to work with the reduced one dimensional Lagrangian \re{YM}-\re{CS}. 
 
Finding solutions within the general YM Ansatz \re{a02}, which after setting $\vec A_r=0$ still
features six independent functions, is technically a difficult task. 
A further consistent trucation of the general Ansatz is $\f^2=\chi^2=0$, leading to an EYMCS system with six unknown
functions, four of them being gauge potentials parametrising the gauge field, and, two metric functions.  Indeed, the
two gauge functions suppressed are redundent and would only be excited in an eventual stability analysis of our
$sphalerons$. 

To make connection with   notations used in  previous   work ~\cite{Okuyama:2002mh,Volkov:2001tb,Brihaye:2002hr}
on $d=5$ EYM solutions, we  adopt the notation 
\begin{eqnarray}
\phi^1(r)=\tilde w(r),~~\phi^3(r)=w(r),~~
\chi^1(r)=\tilde V(r),~~\chi^3(r)=V(r).
\end{eqnarray}

The resulting system has some residual symmetry under a rotation of the 'doublets'
$w(r),\tilde w(r)$ and $V(r),\tilde V(r)$ with the same constant angle
$u$ ($e.g.$ $w\to w \cos u+\tilde w \sin u$ etc.)
One can use this symmetry to consistently set $\tilde w(r)=\tilde V(r)=0$
(or $ w(r)= V(r)=0$) which results in a $particular \ truncation$ of the system, which we shall exploit
in Section 3.2.
Note  that for configurations with $\tilde w(r) = \tilde V(r) =0$
the gauge potentials are invariant under the "chiral" transformations
generated by $\Sigma_5$. The configurations with $w(r)=V(r)=0$ instead change
just by a sign under the same transformations.
Also, this Ansatz is invariant under the parity   reflections  transformation $\phi^a \to -\phi^a,~\chi^a \to -\chi^a $.
The asymmetry beween $(w,V)$ and  $(\tilde w, \tilde V)$ is manifested by the different set of boundary conditions
they satisfy.

\subsection{The equations and boundary conditions }
Inserting this Ansatz into the
action (\ref{action}), the EYMCS field equations (\ref{Einstein-eqs}) reduce to
(to simplify the notation we denote
$\alpha^2=16\pi G/(3e^2)$ and absorb a factor of $1/e$ in the expression of $\kappa$):
\begin{eqnarray}
\nonumber
&&m'=\frac{1}{2}\alpha^2 
\left(
3r \bigg(
N(w'^2+\tilde w'^2)
+\frac{(w^2+\tilde w^2-1)^2}{r^2}
\bigg)
+\frac{r^3}{\sigma^2}\left( V'^2+\tilde V'^2+\frac{3}{r^2N}(\tilde V w-V \tilde w)^2 \right)
\right),
\\
\label{eqs}
&&\frac{\sigma'}{\sigma}=\frac{3\alpha^2}{2r}
\left(
w'^2+\tilde w'^2+\frac{1}{N^2\sigma^2}(\tilde V w-V \tilde w)^2
\right),
\\
\nonumber
&&(r\sigma Nw')'=r\sigma 
\left(
\frac{2w(w^2+\tilde w^2-1)}{r^2}
+\frac{\tilde V}{\sigma^2 N}(V \tilde w-\tilde V w)
\right)
+4 \kappa 
\left(
 V'(w^2+\tilde w^2-1) +2\tilde w'(V \tilde  w-\tilde V  w)
 \right),
 \\
 \nonumber
&& (r\sigma N \tilde w')'=r\sigma 
\left(
\frac{2\tilde  w(w^2+\tilde w^2-1)}{r^2}
+\frac{  V}{\sigma^2 N}(\tilde V  w- V \tilde w)
\right)
+4 \kappa 
\left(
 \tilde V'(w^2+\tilde w^2-1) +2  w'(\tilde V   w- V  \tilde w)
 \right),
 \\
 \nonumber
 &&\big (\frac{r^3 V'}{\sigma}\big )'=\frac{3r}{\sigma N}\tilde w ( V \tilde w-  \tilde V  w)
 +12 \kappa (w^2+\tilde w^2-1)w',
 \\
  \nonumber
  &&\big (\frac{r^3 \tilde  V'}{\sigma}\big )'=\frac{3r}{\sigma N}  w ( \tilde V  w-   V  \tilde w)
 +12 \kappa (w^2+\tilde w^2-1)\tilde w',
\end{eqnarray}
together with the constraint equation
\begin{eqnarray}
\label{constr1} 
\frac{r^3}{\sigma}(\tilde V V'-V \tilde V')+3r N\sigma (w \tilde w'-\tilde w  w' )
-12 \kappa (\tilde V w-V \tilde w)(w^2+\tilde w^2-1)=0\,,
\end{eqnarray}
which originates from the variational equation for $\delta\vec A_r$ (where a prime denotes a
derivative with respect to $r$).

These equations support both globally regular and black hole
solutions.
  The only known closed form solutions of these equations are discussed in the next subsection 
and are trivial in some sense, since the magnetic gauge potentials do not feature any dependence on $r$.
However, it is concievable that non-Abelian analytic solutions can 
 be found  by studying the first order
Bogomol'nyi equations of the full ${\cal{N}}=8$ gauged supergravity model, with all scalar functions included.  
This was the case of other gauged supergravity theories, the most famous example
being the Chamseddine-Volkov solution \cite{Chamseddine:1997nm} of the
${\cal{N}}=4$, $d=4$ Freedman-Schwarz model \cite{Freedman:1978ra}. One might
therefore expect the full ${\cal{N}}=8$ model to support BPS 
solutions describing also non-Abelian globally regular solitons, which actually constrasts with the 
case of an Abelian truncation. However, given the large number of matter functions, even finding the explicit form of the
first order Bogomol'nyi equations of the  ${\cal{N}}=8$ supergravity model with non-Abelian fields
is bound to be a very difficult task, which has not been addressed so far in the literature. This is not surprising
since the analogous task in the construction of the Chamseddine-Volkov solution involves simply a $SU(2)$ gauge field and
a single dilaton field.

However, one can analyse the  properties of the solutions of the system (\ref{eqs}) by using a combination of
analytic and numerical methods, which is sufficient for most purposes.

The globally regular configurations are nontrivial deformations of the AdS$_5$ 
and have the following expansions near the origin  $r=0$:
\begin{eqnarray}
\nonumber
&&w(r)=1-br^2+O(r^4),
~~\tilde w(r)=\frac{g_1 (V(0)-24 \kappa b \sigma_0)}{9 \sigma_0^2}r^3+O(r^5),
\\
\label{origin}
&&V(r)=V(0)+6 b^2 \kappa \sigma_0 r^2+O(r^4),~~
\tilde V(r)=g_1 r+O(r^3),
\\
\nonumber
&& m(r)= \frac{\alpha^2(g_1^2+6b^2\sigma_0^2)}{2 \sigma_0^2}r^4+O(r^6),~~
\sigma(r)=\sigma_0+\frac{3\alpha^2(g_1^2+4 b^2 \sigma_0^2)}{4\sigma_0}r^2+O(r^4).
\end{eqnarray}
The free parameters are $b=-\frac{1}{2}w''(0),~V(0),~g_1=\tilde V'(0)$ and $\sigma_0=\sigma(0)$.
The coefficients of all higher order terms in the $r\to 0$ expansion are fixed by these parameters.

We are also interested in solutions having a regular event horizon at $r=r_h>0$ and representing non-Abelian
generalisations of the Reissner-Nordstr\"om-AdS$_5$ (RNAdS) black hole. To simplify the general picture we shall
consider mainly nonextremal black holes, in which case $N(r)$ has a single zero at $r=r_h$ and $\sigma(r_h)>0$.
 We expect that extremal black holes also exist for the full $SO(6)$ theory, but we have restricted their
numerical construction only to 
 the $particular\ truncation$ ($\tilde w =\tilde V =0$) of the system, alluded to in the previous subsection, which is
of course, a consistent truncation.  We have made 
this restriction simply due to
our desire to render the numerical task easier. From our study of this $particular$ truncation of the system,
we deduce that it is likely extremal black holes exist also for the full $SO(6)$ theory.

For the nonextremal case, the field equations imply the following behaviour as $r \to r_h$ in terms of five
parameters $w_h=w(r_h), \tilde w_h=\tilde w(r_h)$, $V_h=V(r_h)$, $V_1=V'(r_h)$ and $\sigma_h=\sigma(r_h)$:
\begin{eqnarray}
\nonumber
&&w(r)=w_h+w_1(r-r_h)+O(r-r_h)^2,~~\tilde w(r)=\tilde w_h+ \tilde w_1(r-r_h)+O(r-r_h)^2,
\\
\label{eh}
&&V(r)=V_h+V_1(r-r_h)+O(r-r_h)^2,~~\tilde V(r)=\frac{\tilde w_h V_h}{w_h}+ \frac{\tilde w_h V_1}{w_h}(r-r_h)+O(r-r_h)^2,
\\
\nonumber
&&m(r)=r_h^2(1+\frac{r_h^2}{\ell^2})+
\frac{\alpha^2}{2}
\left(
\frac{r_h^3V_1^2(w_h^2+\tilde w_h^2)}{\sigma_h^2w_h^2} +\frac{3(w_h^2+\tilde w_h^2-1)^2}{r_h} 
\right)
(r-r_h)+O(r-r_h)^2,
\\
\nonumber
&&\sigma(r)=\sigma_h+\sigma_1(r-r_h)+O(r-r_h)^2,
\end{eqnarray}
where
\begin{eqnarray}
\nonumber
&&w_1=-\frac{4r_h\ell^2\sigma_hw_h^2(2\kappa r_h V_1+\sigma_h w_h)(w_h^2+\tilde w_h^2-1)}
{-4r_h^2 (2r_h^2+\ell^2)\sigma_h^2w_h^2+\alpha^2\ell^2 
(3\sigma_h^2w_h^2(w_h^2+\tilde w_h^2-1)^2+r_h^4V_1^2(w_h^2+\tilde w_h^2))},
\\
\label{eh-exp1}
&&\tilde w_1=
-\frac{4r_h\ell^2\sigma_h w_h \tilde w_h(2\kappa r_h V_1+\sigma_h w_h)(w_h^2+\tilde w_h^2-1)}
{-4r_h^2 \ell^2 \sigma_h^2 w_h^2+3 \alpha^2 \ell^2 \sigma_h^2 w_h^2 (w_h^2+\tilde w_h^2-1)^2
+r_h^4(-8 \sigma_h^2w_h^2+\alpha^2\ell^2 V_1^2(w_h^2+\tilde w_h^2))
},
\\
\nonumber
&&\sigma_1=\frac{24 \alpha^2 r_h \ell^4 \sigma_h^3 w_h^2 (2\kappa r_h V_1 +\sigma_h w_h)^2(w_h^2+\tilde w_h^2-1)^2(w_h^2+\tilde w_h^2)}
{
\bigg(
4 r_h^2\ell^2 \sigma_h^2 w_h^2-3 \alpha^2 \ell^2 \sigma_h^2 w_h^2(w_h^2+\tilde w_h^2-1)^2
+r_h^4(8\sigma_h^2 w_h^2-\alpha^2\ell^2 V_1^2(w_h^2+\tilde w_h^2) 
)
\bigg)^2 
}~~.
\end{eqnarray}

\subsubsection{Large $r$ asymptotic expansions}
The expansion at infinity of the  solutions is more complicated  and involves separate analyses in the $generic$
and the $special$ cases.

In the $generic$ case, the potentials parametrising the non-Abelian gauge field 
take arbitrary values as $r\to \infty$, with the leading order behaviour
\begin{eqnarray}
\nonumber
&&w(r)=w_0+\frac{w_2}{r^2}+w_c\frac{ \log r}{r^2}+\dots,~~
\tilde w(r)=\tilde w_0+\frac{\tilde w_2}{r^2} +\tilde w_c\frac{ \log r}{r^2}+\dots,~~
\\
\label{asimpt-div}
&&V(r)=V_0+\frac{q}{r^2}+V_c\frac{ \log r}{r^2}+\dots,~~ 
\tilde V(r)=\tilde V_0+\frac{\tilde q}{r^2}+\tilde V_c\frac{ \log r}{r^2}+\dots,~~ 
\\
\nonumber
&&
m(r)=M_0+\frac{3}{2}\alpha^2 \left((w_0^2+\tilde w_0^2-1)^2+\ell^2(\tilde V_0 w_0-V_0 \tilde  w_0)^2 \right)\log r,~~
\sigma(r)=1-\frac{\alpha^2(w_c^2+\tilde w_c^2)\log^2 r}{r^6}+\dots,~~
\end{eqnarray}
with
\begin{eqnarray}
\nonumber
&&w_c=-\frac{\ell^2}{2}\left (2w_0 (w_0^2+\tilde w_0^2-1)+\tilde V_0 \ell^2 (V_0 \tilde w_0-\tilde V_0  w_0)\right ),
\\
\label{Vc}
&& \tilde w_c=-\frac{\ell^2}{2}\left (2\tilde  w_0 (w_0^2+\tilde w_0^2-1)+  V_0 \ell^2 (\tilde V_0  w_0- V_0 \tilde w_0)\right ),
\\
\nonumber
&&V_c=-\frac{3}{2}\ell^2 \tilde w_0 ( V_0  \tilde w_0- \tilde V_0  w_0),~~
\tilde V_c=-\frac{3}{2}\ell^2  w_0 ( \tilde V_0  w_0- V_0   \tilde w_0),
\end{eqnarray}
where $w_0,\tilde w_0,V_0,\tilde V_0$ and $w_2,\tilde w_2,q,\tilde q$ are arbitrary parameters satisfying the
constraint 
\begin{eqnarray}
3(w_2 \tilde w_0-\tilde w_2  w_0)+\ell^2(\tilde q V_0- q  \tilde V_0)
+6 \ell^2 \kappa ( V_0 \tilde w_0-  \tilde  V_0 w_0)(w_0^2+\tilde w_0^2-1)=0.
\end{eqnarray}
Thus, similar to the well known case
of a SO(3) gauge group  \cite{Okuyama:2002mh}, the generic non-Abelian 
configurations have a nonvanishing magnetic field on the AdS boundary ($i.e.$ $F_{\mu\nu}F^{\mu\nu}|_{r\to \infty}\neq 0$).
As a result, one can see from the above relations that the mass function $m(r)$,  and hence also the action of these
solutions, diverge logarithmically~\footnote{The existence of a logarithmic divergence in the action
is a known property of some classes of AdS$_5$ solutions which are endowed with
with a special boundary geometry  \cite{Emparan:1999pm}. 
The coefficients of the divergent terms there are related to the conformal 
Weyl anomaly in the dual theory \cite{Skenderis:2000in,odintsov}.
However, this is not the case for the non-Abelian AdS$_5$ configurations here, which have the same boundary metric
as the Schwarzschild-AdS  solution and hence feature no Weyl anomaly in the dual CFT.}.

However, despite the divergence of the mass, the spacetime is  still asymptotically AdS, 
the large $r$ behaviour  of the metric function $N(r)$ being $N(r)\to r^2/\ell^2+1$.
Asymptotically AdS solutions with 
diverging mass have been considered recently by some authors, mainly for a scalar field in the bulk
(see e.g. \cite{Hertog:2004dr}). In this case it might be possible to relax the standard asymptotic
conditions without loosing the original symmetries, but modifying the charges in order to 
take into account the presence of matter fields. In Section 5 of this work
we shall argue  that this is also the case of the EYMCS configurations with the general asymptotics (\ref{asimpt-div}).
By using a counterterm approach,  one can define a mass for  these solutions,
which is fixed by the parameter $M_0$ appearing in (\ref{asimpt-div}).
(Note that for generic solutions not only $m(r)$ diverges logarithmically as $r\to\infty$ but also 
the terms $r^3 V'(r)$ and $r^3 \tilde V'(r)$ which, as argued in Section 4, fixes the electric
charge(s) of the solutions. In the numerics, we have studied mainly the solutions with $V_c=\tilde V_c=0$.)

 For the $special$ configurations, which support finite mass, the required asymptotic behaviour at large $r$ is
$|\f^a_0|\to 1$ and $ |\vec \f \times \vec \chi | \to 0$
($i.e.$ $w_0^2+\tilde w_0^2 \to 1$ and 
$w_0\tilde V_0-\tilde w_0 V_0 \to 0$).  
 These conditions can be satisfied $only$ in the presence of the CS term. Different from the case of a simple
EYM theory \cite{Okuyama:2002mh}, the existence 
of  finite mass configurations here is not forbidden by the Derrick-type scaling argument.
Indeed, the numerics in the following Section indicate the existence of a subset solutions with finite mass,
with the following expansion at infinity  
\begin{eqnarray}
\nonumber
&&w(r)=\sin \alpha+\frac{w_2}{r^2}+O(1/r^4),~~
\tilde w(r)=\cos \alpha+\frac{\tilde w_2}{r^2}+O(1/r^4),~~
\\
\label{infinity}
&&V(r)=\Phi \sin \alpha+\frac{q}{r^2}+O(1/r^4),~~
\tilde V(r)=\Phi \cos \alpha+\frac{\tilde q}{r^2}+O(1/r^4),~~
\\
\nonumber
&&
m(r)=M-\frac{\alpha^2(q^2+\tilde q^2)\ell^2+3(w_2^2+\tilde w_2^2))}{r^2\ell^2}+O(1/r^4),~~
\sigma(r)=1-\frac{\alpha^2(w_2^2+\tilde w_2^2)}{r^6}+O(1/r^8),~~
\end{eqnarray}
where the amplitude of the electric potential at infinity is fixed by
\begin{eqnarray}
\Phi=\frac{3(w_2 \cos \alpha -\tilde w_2 \sin\alpha)}{\ell^2(q \cos \alpha -\tilde q \sin\alpha)}.
\end{eqnarray}
Thus the free parameters in the far field expansion of this $special$ set of solutions are $M_0$, 
$\alpha=\arctan(V(\infty)/\tilde V(\infty))$ and the coefficients $q,\tilde q$, $w_2,\tilde w_2$
of the $1/r^2$ decaying terms  in the non-Abelian potentials.

\subsection{Particular cases}
The simplest solution of the field equations has pure 
gauge fields ($F_{\mu \nu}=0$) and corresponds to the
Schwarzschild-AdS$_5$  black hole   
\begin{eqnarray}
\label{ex-sol2}
N(r) =1+\frac{r^2}{\ell^2}-\frac {M}{r^2}~,
~~
\sigma(r)=1~,~~w(r)=~\sin \alpha,~~\tilde w(r) = \cos \alpha,~~V(r)= \Phi \sin \alpha,~~\tilde V(r)=\Phi \cos \alpha,
\end{eqnarray}
 where $\Phi,\alpha$ are arbitrary constant.

The embedding of the RNAdS Abelian solution
 is recovered for  
\begin{eqnarray}
\label{ex-sol3}
N(r) =1+\frac{r^2}{\ell^2}-\frac {M }{r^2}+\frac{\alpha^2 q^2}{r^4}~,
~~
\sigma(r)=1,~~\tilde w(r)=\tilde V(r)=0,~~w(r) =\pm 1,~~V(r)=\Phi+\frac{q}{r^2}.
\end{eqnarray} 
 The only AdS exact solutions  with nontrivial non-Abelian fields known so far is:
\begin{eqnarray}
\label{ex-sol}
N(r) =1+\frac{r^2}{\ell^2}-\frac {M_0+\frac{3\alpha^2}{2}\log r}{r^2}~,
~~
\sigma(r)=1~,~~w(r)=\tilde w(r)=0,~~ V(r)=\tilde V(r)=0,
\end{eqnarray}
(with $M_0$ an arbitrary positive constant).
This solution was obtained in  \cite{Okuyama:2002mh} and describes
a Reissner-Nordstr\"om type geometry in EYM theory
with a gauge group $SO(3)$ (note that the mass function is logarithmically divergent in this case).
Its embedding in the $SO(6)$ gauged supergravity model and the extremal limit 
has been discussed in the recent work \cite{Cvetic:2009id}.

A particularly interesting model is found by taking  
$\tilde w(r)=\tilde V(r)=0$ (or equivalently  $w(r)=V(r)=0$). 
 This is our $particular\ truncation$ of the full $SO(6)$ model. The resulting solutions are those of the
$SU(2)\times U(1)$ truncation of the model\footnote{This truncation of the full $SO(6)$ model
shares a number of common features with the five dimensional
$N=4$ gauged $SU(2)\times U(1)$ supergravity model considered in \cite{Chamseddine:2001hk}.
For example, a first integral similar to  (\ref{1SU2U1}) appears there also.
However, the solutions in \cite{Chamseddine:2001hk} have an extra dilaton field with a Liouville 
potential and thus are not asymptotically AdS.  } parametrised in terms of the representations of the algebra of $SU(4)$
instead of $SO(6)$. (Of course in that case both gauge groups have the same gauge coupling constant.) 
One can see that the CS term is still nontrivial in this case
\begin{eqnarray}
&&{\cal L}_{\rm{CS}} =i~ \kappa \,
V(r)\vep_{a_1a_2 a_3 a_4}\mbox{Tr}\, \big \{
 F^{a_1 a_2}F^{a_3 a_4}  \big \}=12 \kappa V(r) w'(r)(w^2(r)-1),
\label{CS-SU2U1}
\end{eqnarray}
(with $a_i=1,\dots 4$).
As we shall argue in the next Section,   the solutions  of this particular truncation 
contain already the basic features of the full model.
However, they are much easier to study  numerically.

The YM-CS equations in this case admit the first integral
 \begin{eqnarray}
\label{1SU2U1}
V'(r)=\frac{\sigma}{r^3}\big(K+4\kappa w(w^2-3)\big), 
\end{eqnarray}
with $K$ being an integration constant.
One can easily see that the solutions are regular at the origin, $r = 0$, only if $K = 8\kappa$.
The value of $K$ is not fixed $a\ priori$ for black hole solutions.

\begin{figure}[ht]
  
 \hbox to\linewidth{\hss%
	\resizebox{8cm}{6cm}{\includegraphics{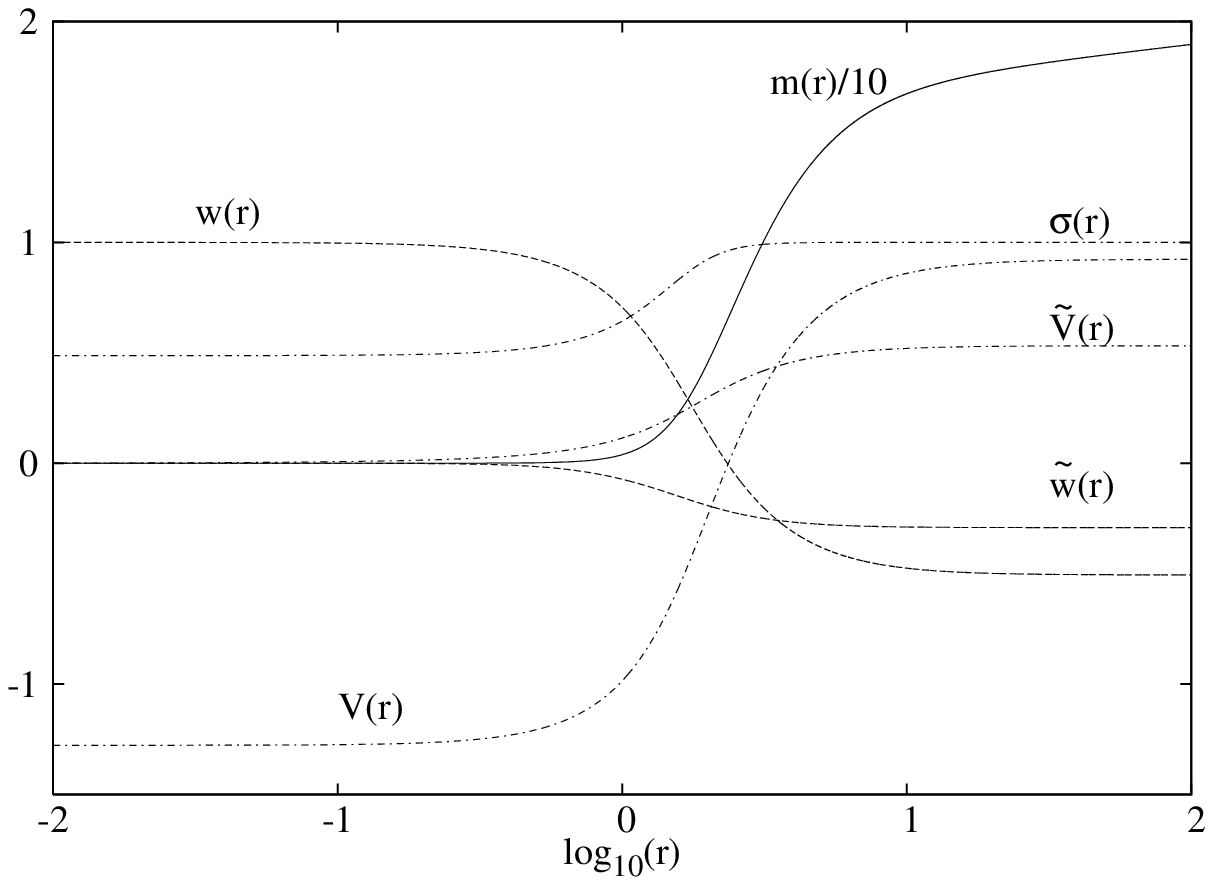}}
\hspace{5mm}%
        \resizebox{8cm}{6cm}{\includegraphics{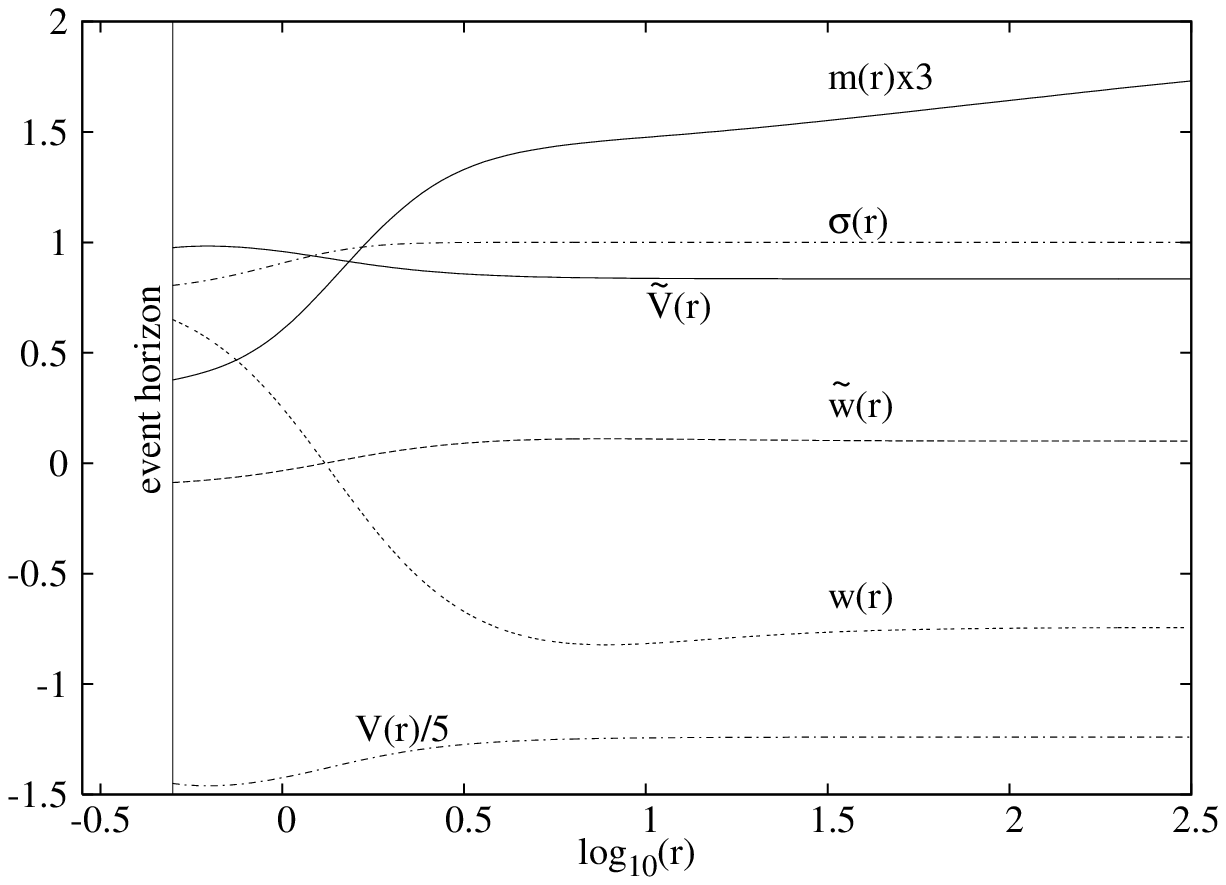}}	
\hss}
 
\caption{
{\small
The profiles of a typical globally regular solution (left) and a black hole
solution (right) of the EYMCS equations are presented as a function of the radial coordinate $r$. 
For these solutions, the mass function $m(r)$ diverges logarithmically as $r\to \infty$.
}
 }
\label{fig1}
\end{figure}

The asymptotics of the $SU(2)\times U(1)$ solutions can easily be read 
from the general relations (\ref{origin}), (\ref{asimpt-div}) and (\ref{infinity}).
At infinity, the finite mass solutions have $\alpha=\pm \pi/2$ in (\ref{infinity}), with the gauge potentials
\begin{eqnarray}
\label{FI}
V(r)=\Phi-\frac{ 8\kappa \mp K  }{2r^2}+O(1/r^4), ~~~w(r)=\pm 1+\frac{w_2}{r^2}+O(1/r^4).
\end{eqnarray} 
  Thus, for the case  $w(\infty)=-1$ studied in this work, 
the parameter  $q$  fixing the Abelian electric charge of these solutions
is $q=-(4\kappa + K/2) $ for black holes and $q=-8 \kappa$ for globally regular solutions. 
  
 \section{Numerical results}
We start by noticing that the equations (\ref{eqs}) are not affected by the transformation:
\begin{eqnarray}
\label{ss1}
 r\to \lambda r, ~~m\to \lambda^2 m,~~\ell\to\lambda \ell,~~V\to V/\lambda,~~\tilde V\to \tilde V/\lambda,~~\alpha\to \lambda \alpha~,
 ~~\kappa\to \lambda \kappa,
\end{eqnarray}
while $w,\tilde w$ and $\sigma$ remain unchanged. It follows that one can always take an arbitrary positive value for
$\alpha$. The usual choice is $\alpha=1$, which fixes  the EYM length scale
$L=\sqrt{8\pi G/(3 e^2) }$, while the mass scale is fixed by ${\cal M}=8\pi/(3 e^2)$.
All other quantities get multiplied with suitable factors of $L$.
However, in this Section, to avoid cluttering our expressions with a complicated dependence on $(G,e)$,
we fix the value of $\alpha$ at $\alpha=1$,
and ignore the extra factors of $e$ and $G$ in the expressions of various global quantities.

Therefore the remaining input parameters are the AdS length scale $\ell$ and the CS coupling constant $\kappa$.
Determining the pattern  of the solutions in the parameter space represents a very complex task which is outside
the scope of this paper. Instead, we analysed in detail a few particular classes of solutions which, hopefully,
reflect all relevant properties of the general pattern. For definiteness we set 
$\ell=1$ in our numerical analysis, although we have found nontrivial 
solutions also for other values of the cosmological constant\footnote{In particular, 
the finite energy solutions
survive in the limit $\Lambda \to 0$, being supported by the CS term (this
contrasts strongly with the case of a pure EYM theory \cite{Volkov:2001tb}). A discussion of the 
asymptotically flat EYMCS solutions will be presented elsewhere.}.

\begin{figure}[ht]
\hbox to\linewidth{\hss%
	\resizebox{8cm}{6cm}{\includegraphics{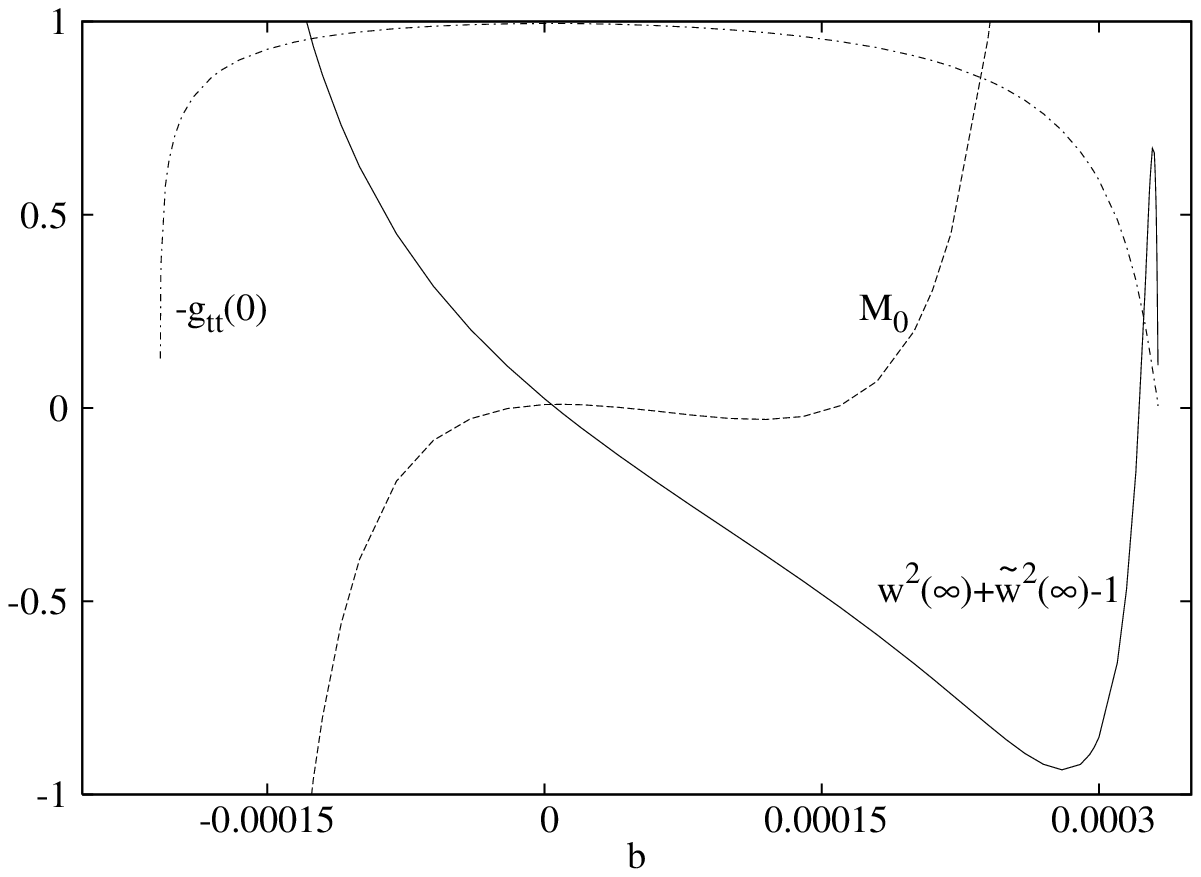}}
\hspace{5mm}%
        \resizebox{8cm}{6cm}{\includegraphics{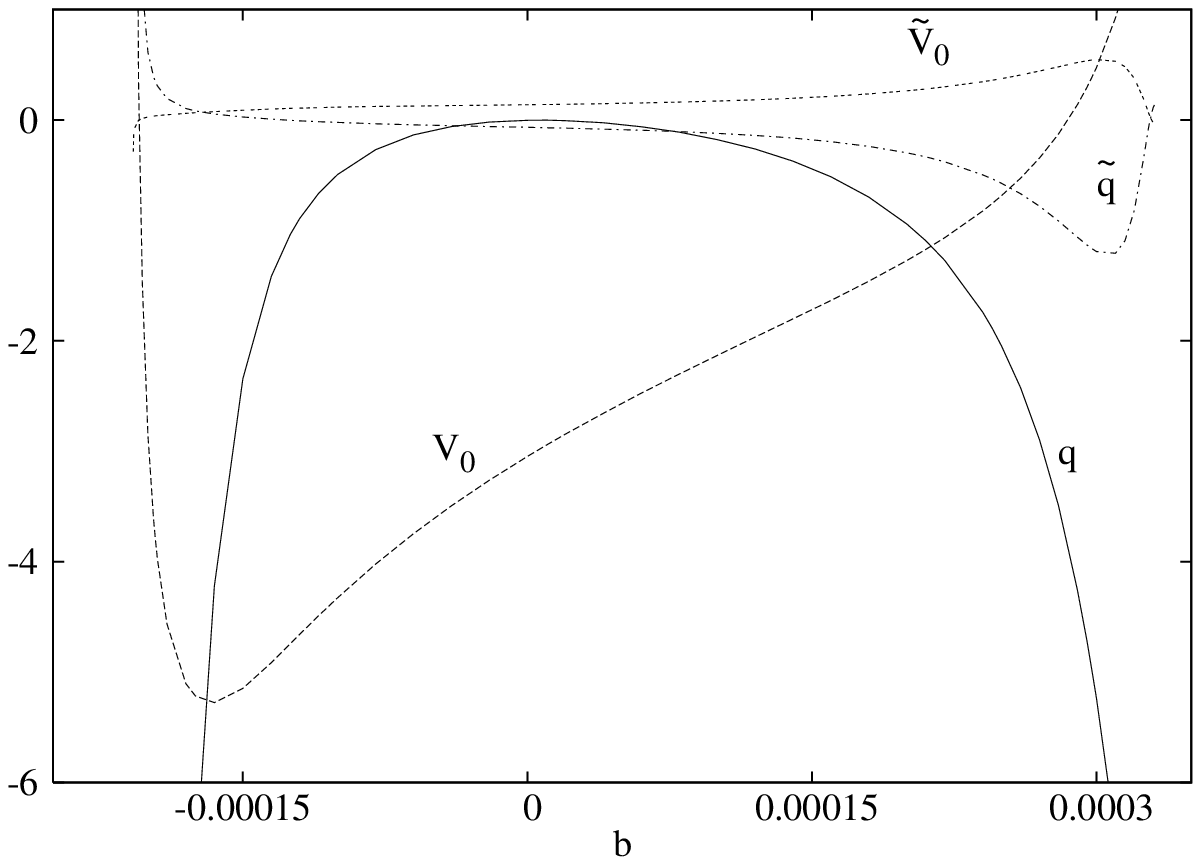}}	
\hss}

\caption{
{\small
A number of relevant parameters are plotted as functions of the coefficient $b$ for 
globally regular solutions of the EYMCS model with $\ell=1$, $\kappa=1$.
These solutions have finite electric charges, which are fixed by $q$ and $\tilde q$.
 }
 }
\label{fig42}
\end{figure}

Instead we have looked for the dependence of the solutions
on the value of the CS  coefficient $\kappa$, which has not been 
fixed $a\ priori$. (This investigation has been partly motivated by the study in \cite{Kunz:2006xk}
of the Einstein-Maxwell-CS system, which revealed a nontrivial dependence of the properties of the solutions   
on the value of $\kappa$. We shall see that this is also the case for the solutions constructed in this work,
which feature a critical value of CS coefficient.)

The resulting set of six ordinary differential equations\footnote{Although we have solved the second order
YM equations in (\ref{eqs}), we have also monitored
the constraint (\ref{constr1}), which was always satisfied with very good accuracy.
Also, the  equation (\ref{constr1}) has  been used to construct the asymptotic expansions
(\ref{origin}), (\ref{eh}), (\ref{asimpt-div}) and (\ref{infinity}).
}
is solved with suitable boundary conditions which result from (\ref{origin}), (\ref{eh}), (\ref{asimpt-div}) and
(\ref{infinity}). The numerics employs a collocation method for boundary-value ordinary
differential equations equipped with an adaptive mesh selection procedure \cite{colsys}.
Typical mesh sizes include $10^3-10^4$ points. The solutions have a relative accuracy of $10^{-7}$.
In addition to employing this algorithm, some solutions were also constructed by  using a standard Runge-Kutta
ordinary  differential equation solver. In this approach we 
evaluate the initial conditions at $r=10^{-5}$ (or $r=r_h+10^{-5})$, for global tolerance $10^{-12}$,
adjusting  for shooting parameters and integrating  towards  $r\to\infty$.
We have confirmed that there is good agreement between the results obtained with these two different methods.

The properties of the solutions depend on the input parameters, but it is rather difficult to find a
general pattern. However, a feature shared by all asymptotically AdS solutions is that the metric functions $m(r)$, $\sigma(r)$
monotonically approach their asymptotic values, which can easily be seen from the corresponding field equations.
Also, so far we could not find solutions where the electric potentials $V(r)$, $\tilde V(r)$
present oscillations, even though such solutions are allowed.

 \subsection{The generic  solutions }
 
The generic solutions studied here are for the $d=4+1$ dimensional
model, which features a Chern--Simons (CS) term. But solutions
with similar properties are found also in the EYM model with no CS term. All these
solutions bear a qualitative similarity to those of the
more familiar $d=3+1$ EYM model \cite{Winstanley:1998sn}, \cite{Bjoraker:2000qd}. 
 These solutions can also be seen as higher gauge group generalisations of the
EYM solutions in \cite{Okuyama:2002mh}, but unlike the latter they feature a nontrivial electric potential,
made possible by the larger gauge group. (The electric potential
necessarily vanishes for a $d=5$  static, spherically symmetric  $SU(2)$ gauge field). 

Considering first the case of globally regular configurations, one finds that
solutions approaching asymptotically the AdS$_5$ background
exist for compact intervals of the initial parameters $w''(0),~V(0),~V'(0)$ and $\sigma(0)$.
 The values of the parameters $w_0,\tilde w_0,V_0,\tilde V_0$, $w_2,\tilde w_2$ 
 and $q$ which enter the asymptotics
 %
\begin{figure}[ht]
\hbox to\linewidth{\hss%
	\resizebox{8cm}{6cm}{\includegraphics{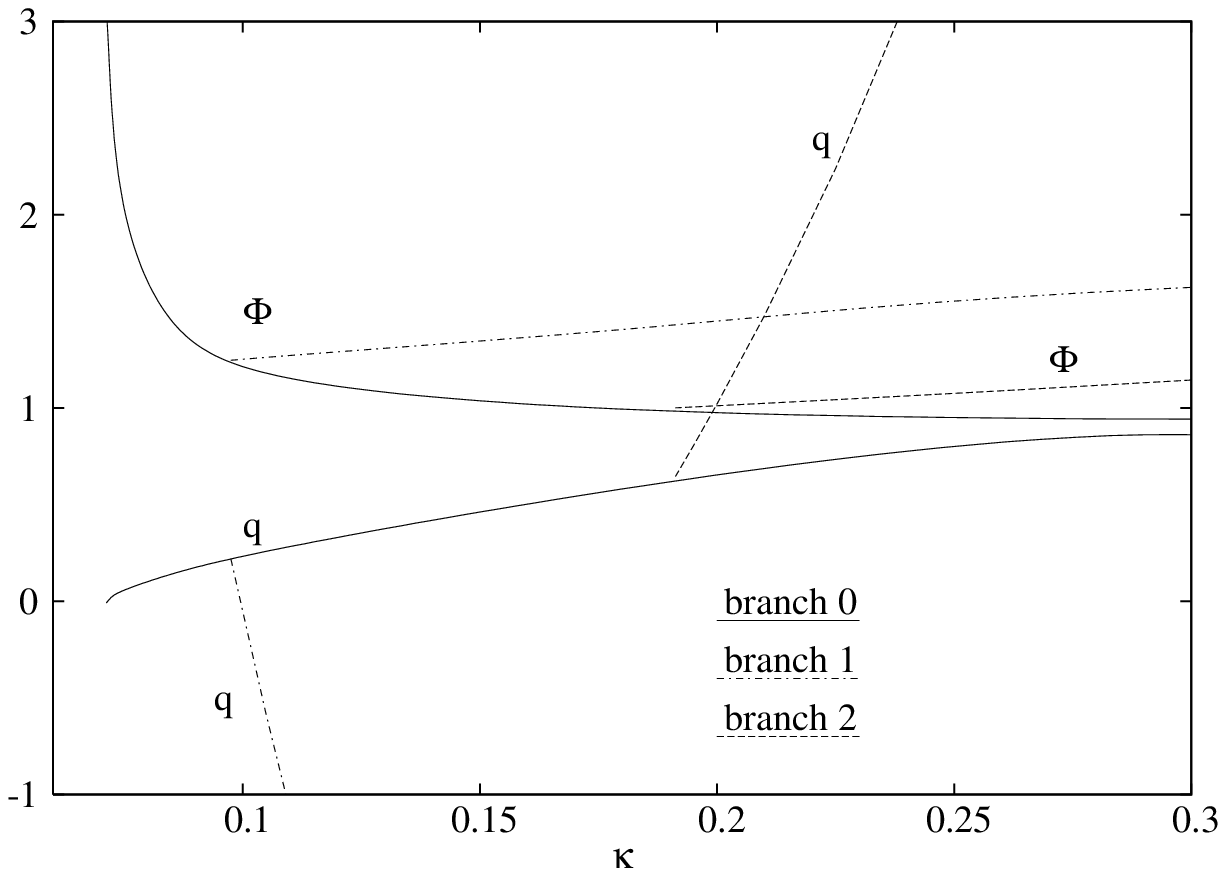}}
\hspace{5mm}%
        \resizebox{8cm}{6cm}{\includegraphics{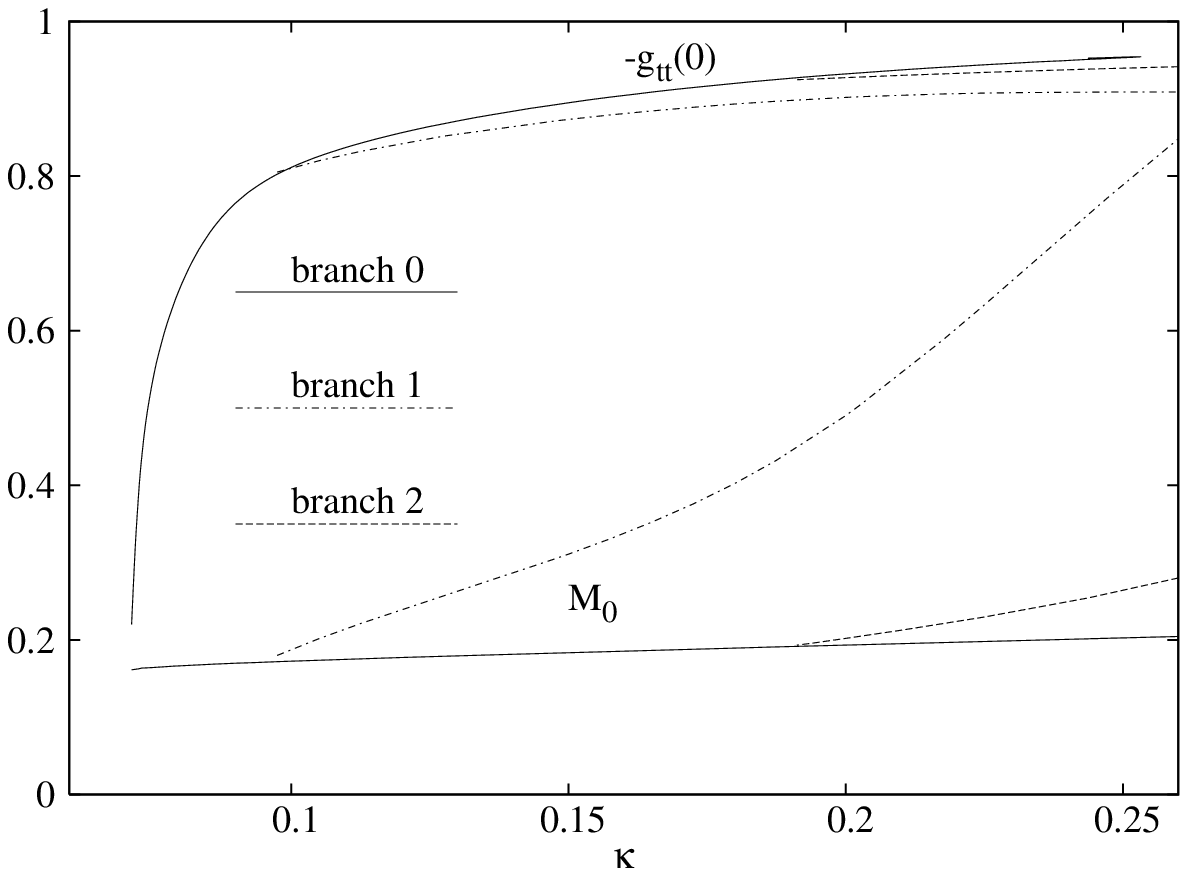}}	
\hss}

\caption{
{\small
A number of relevant parameters are plotted as functions of the Chern-Simons coupling $\kappa$
for finite mass, globally regular solutions of the EYMCS model. 
One can see that new branches of solutions emerge as $\kappa$ increases.
}
}
\label{fig43}
\end{figure}
%
of the solutions are fixed by the numerics.
(There are also branches of solutions with a different asymptotic behaviour, which
stop to exist for finite values of $r$. To study these configurations, one needs to employ
a metric Ansatz different from (\ref{metric-gen}). Such solutions, being not asymptotically AdS,
are of no interest here.)

Of the full set of solutions with AdS asymptotics, we have paid special attention
to the physically more interesting case of configurations with a $1/r^2$ decay of the electric potentials at infinity
($i.e.$ $V_c=\tilde V_c=0$ and $\tilde V_0 w_0-V_0 \tilde w_0=0$), this being the only case reported in this Section. 
These solutions have a finite electric charge, although their mass functions will diverge asymptotically since
$|\vec\phi|=\sqrt{w_0^2+\tilde w_0^2} \not\to 1$ here. 

A typical configuration  with a regular origin is presented in Figure 1 (left), for $\kappa=1$.
One can see that the mass function diverges logarithmically while
$\sigma(r)$, $w(r),\tilde w(r)$ and $V(r),\tilde V(r)$ asymptotically approach some finite values.
Solutions with nodes in $w(r)$, $\tilde w(r)$ were also found.

In Figure 2 we plot a number of relevant parameters
as a function of the coefficient $b$ in the initial data at $r=0$ ($b=-w''(0)/2$),
for a family of  asymptotically AdS solutions. (One of the parameters there is $M_0$ appearing in (\ref{asimpt-div}),
which in Section 5 we argue that it can be taken as the renormalised
mass of the solutions; note that $M_0$ may take also negative values). 
This branch ends for some finite values of $b$, where $-g_{tt}(0)=\sigma(0) \to 0$ while $V_0,q$ diverge. 
The condition $ |\vec \f \times \vec \chi | \to 0$ as $r \to \infty$ has been enforced
by treating $V(0)$ as a shooting parameter. 
Then $\tilde V'(0)$ is a free parameter while $\sigma(0)$ results from the numerics
(the solutions in Figure 2 have $ V'(0)/\sigma_0=0.15$).

The results in Figure 2 show that the for generic solutions
$w^2(\infty)+\tilde w^2(\infty)\not\to 1$. From (\ref{asimpt-div}), this
leads to a divergent mass-energy as defined in the usual way. 
However, one can see that the condition $|\vec\phi|\to 1$ is satisfied for a discrete set of the parameter 
$b$ ($e.g.$ $b\simeq 0.3219 \times 10^{-3}$ and $b\simeq 0.6125 \times 10^{-5}$ for the data in Figure 2).
This suggests the existence of several branches of finite mass solutions parametrised by 
$\tilde V'(0)$ (or, equivalently, $V(0)$), which is confirmed by the results in the next subsection.

Black hole solutions have been found as well, presenting the same general
features. Here also we find a continuum of solutions with arbitrary
values of gauge potentials at infinity, the relevant parameters being the values of the gauge potentials
at the event horizon as given by (\ref{eh}).
Again, finite mass black holes are found only for specific values of the gauge potentials on the horizon.
 
As a general remark, we note that the presence of the CS term is not crucial for the existence of the generic solutions
($i.e.$ with a divergent  mass). We have found solutions with rather similar properties also for $\kappa=0$. Thus, the role of the
CS term is indispensable only  for the construction of $special$, finite mass, solutions to be presented in the next subsection.
 
%
\begin{figure}[ht]
\hbox to\linewidth{\hss%
	\resizebox{8cm}{6cm}{\includegraphics{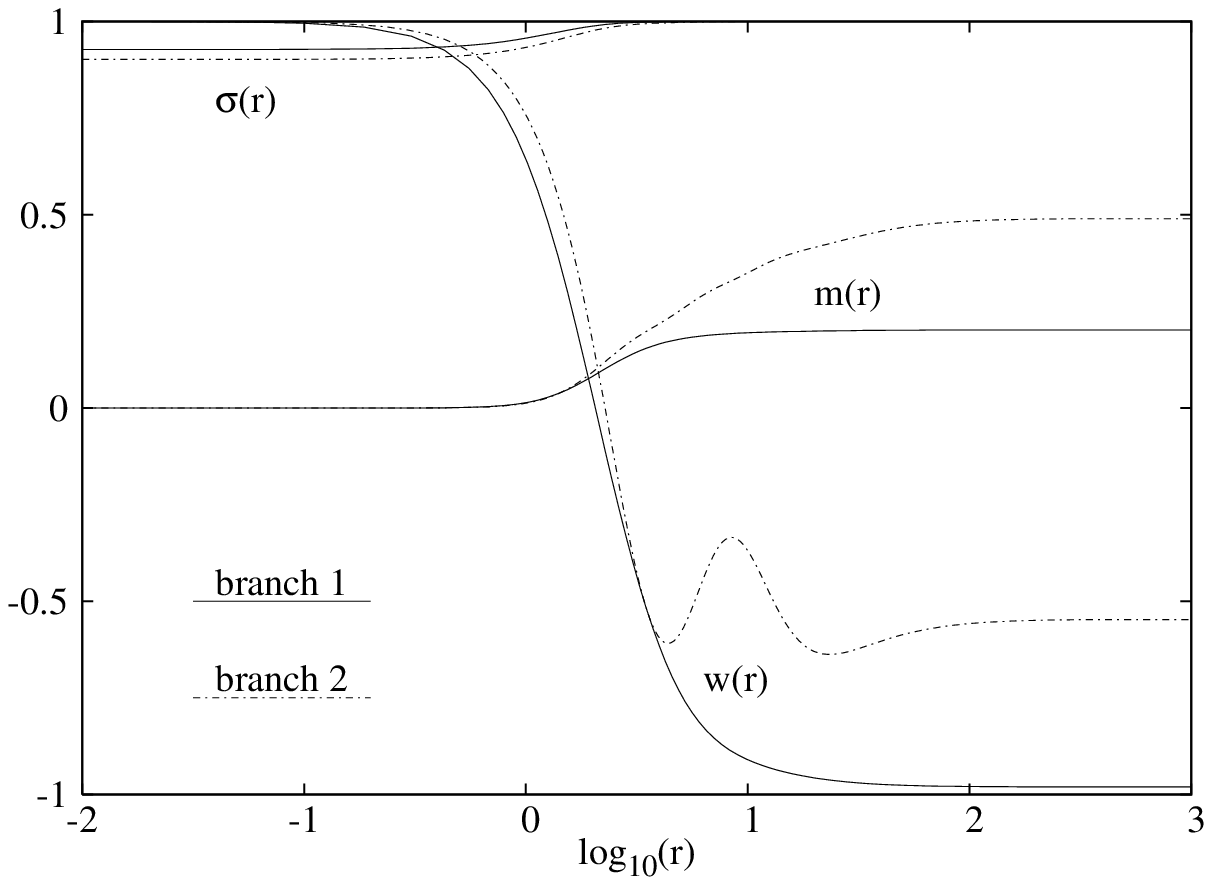}}
\hspace{5mm}%
        \resizebox{8cm}{6cm}{\includegraphics{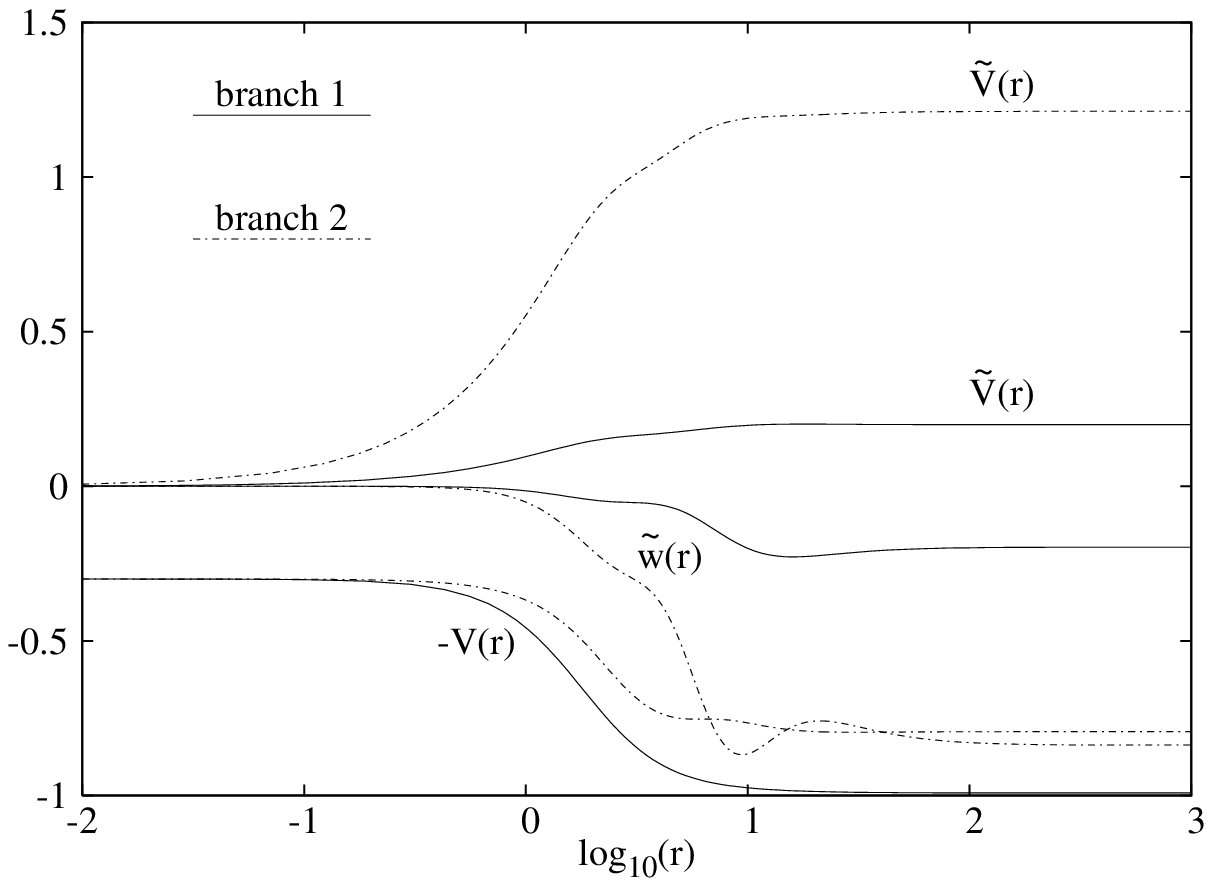}}	
\hss}
\caption{
{\small
The profiles of two typical globally regular EYMS solutions are presented as function of the radial coordinate $r$. 
$m(r)$ and $\sigma(r)$ are metric functions, which $w(r),\tilde w(r)$
and $V(r),\tilde V(r)$ are non-Abelian potentials.
}
 }
\label{fig44}
\end{figure}
\begin{figure}[ht]
\hbox to\linewidth{\hss%
	\resizebox{8cm}{6cm}{\includegraphics{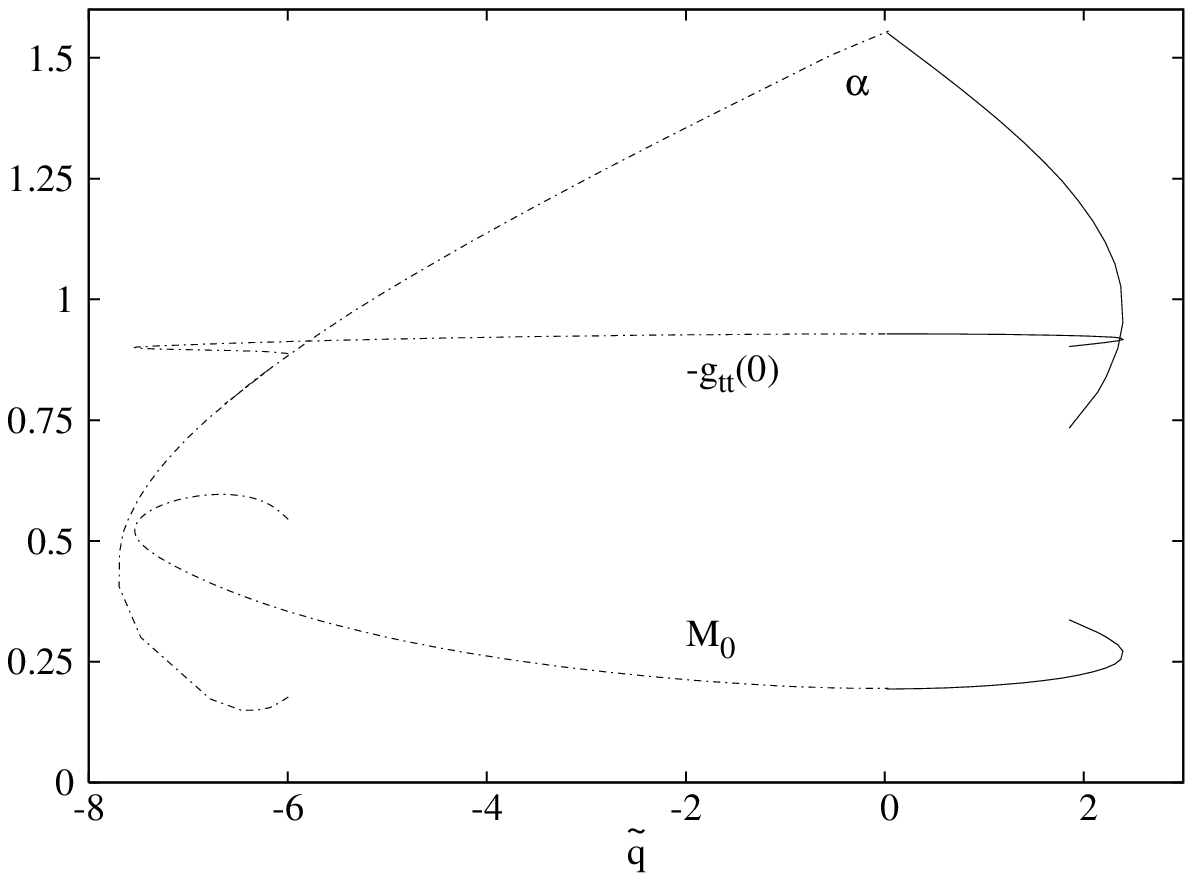}}
\hspace{5mm}%
        \resizebox{8cm}{6cm}{\includegraphics{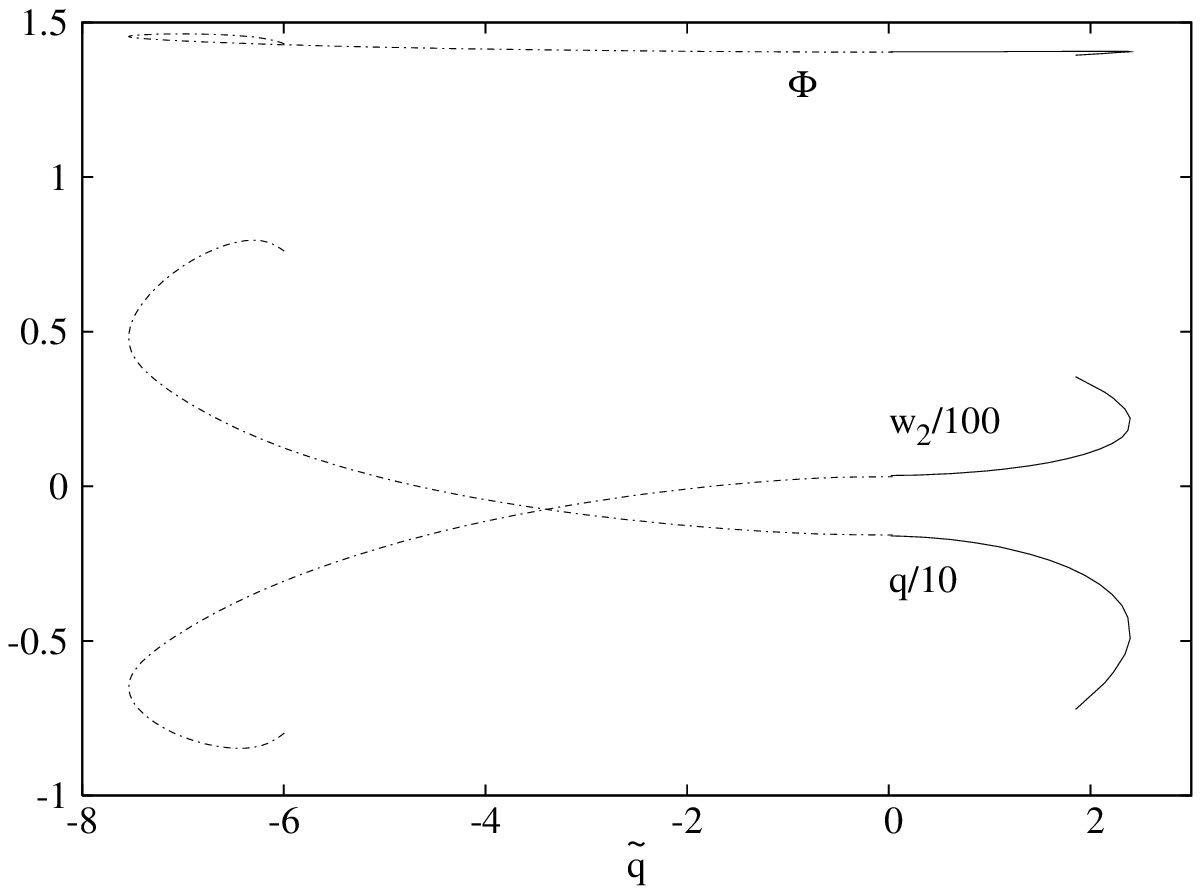}}	
\hss}

\caption{
{\small
A number of relevant parameters are plotted as functions of the electric charge
$\tilde q$ for finite mass, globally regular solutions of the EYMCS model with a
Chern-Simons coefficient $\kappa=0.2$.
 The dashed and solid curves denote different branches of solutions.
}
 }
\label{fig45}
\end{figure}
%
 \subsection{Finite mass solutions}

\subsubsection{Regular configurations}

In the numerics, special attention has been paid to solutions with a finite mass.
As noted above, in this case, three of the four parameters in the data at $r=0$ are fixed.
The remaining free parameter was chosen to be $V(0)$  (similar results are found when $\tilde V(0)$ is chosen instead).

Remembering the invariance under the parity reflection $\vec \phi \to -\vec \phi $, $\vec \chi  \to - \vec \chi $
it turns out to be sufficient to consider $ V(0) \geq 0$ (or $\tilde V(0) \geq 0$). 
 
Thus, corresponding to a choice of the coupling constant $\kappa$ and of the cosmological constant $\Lambda$,
there exist in principle a family of finite mass charged EYMCS solutions labeled by the value at the origin of
one of the electric potentials, in this case, $V(0)$. The pattern of these solutions turns out to be extremely  rich, with some
unexpected features. 
 
In order to illustrate this, we first fix $V(0)=0.3$ and study the solutions as functions of the CS parameter $\kappa$.
(Qualitatively, the same results have been found when considering other values of $V(0)$). 
The numerical results show that a branch of solutions with $\tilde w(r)=\tilde V(r)=0$ always exists for sufficiently
large values of $\kappa$, for instance, $\kappa \geq \kappa_0 \simeq 0.07$ in the present case.
These are the solutions of the reduced $SU(2)\times U(1)$ model with the asymptotic angle $\alpha=\pi/2$.
For convenience, we will refer to this branch as the $main\ branch$.
In the limit $\kappa \to \kappa_0$ the metric function $\sigma(r)$ vanishes at the origin and the solution becomes singular.
For all values of the other parameters, no finite mass solutions have been found for $\kappa < \kappa_0$.

The interesting feature is that new branches of solutions with notrivial functions $\tilde w(r),~\tilde V(r) $
emerge from the main branch at critical values of $\kappa$. In our cases, the first branch of excited solution
appears $\kappa \simeq 0.0953$ and a second branch at $\kappa \simeq 0.1875$.
This behaviour is illustrated in Figure 3, where a number of relavant global parameters are plotted 
a functions of the CS coupling constant $\kappa$. It should be noted that, for a fixed $\kappa$, the
excited solutions have larger masses than the solutions on the $main\ branch$.  

The profiles of the two first excited solutions corresponding to $\kappa=0.2$ 
is presented in Figure 4. One can see that the functions $w,\tilde w$ parametrising the magnetic field
of the excited solutions develop more pronounced oscillations before
becoming constant in the asymptotic region. 
  
It is also natural to study the spectrum of solutions in terms of one of the charges, say $\tilde q$ for  
a fixed value of $\kappa$. The second electric charge $q$ is determined from numerics.
The results of our analysis for the value $\kappa=0.2$ are summarised in Figure 5. 
In this case, two excited solutions are available.
Several relevant parameters  are plotted there
$versus$ $\tilde q$
(see Eqn. \re{infinity}) for two excited solutions.
Fixing the parity symmetry by means of $\tilde V'(0) \geq 0$, the numerical analysis reveals that
the solutions  of the branch "1" (respectively "2") are characterised by negative (respectively positive)
values of $\tilde q$. Also, they exist up to a minimal (respectively maximal) value, say $\tilde q = \tilde q_{cr}$.
For $\kappa=0.2$  we have 
$\tilde q_{cr} \approx -7.8$ and  $\tilde q_{cr} \approx 2.5$ respectively, for the first and second branches.  
In the limit $\tilde q_{cr} \to 0$, the excited  solutions converge to the main solution. The ending of the
branches at $\tilde q= \tilde q_{cr}$  is more subtle. Indeed, our numerical results show that another
family of solutions (with larger mass) exists in the region $|\tilde q | < |\tilde q_{cr}|$, backbending from
the branch coming directly from the main solution. These new branches are shown on Figure 5;
however, we have not attempted to construct further branches in this region, although they are likely exist.

\subsubsection{Black holes}
The EYMCS system presents also black hole solutions which were constructed using similar techniques.
In contrast with the regular solutions presented above, the main thrust here
is confined to the  solutions of our $particular\ truncation$ of the full $SO(6)$ model. 
This restriction is made to simplify an otherwise very complex numerical task.

The Hawking temperature and the entropy of the black holes are given by
\begin{eqnarray}
\label{THS}
T_H= \frac{\sigma(r_h) N'(r_h)}{4 \pi},~~~~S=\frac{A_H}{4G},~~~~~{\rm with}~~A_H=V_3 r_h^3,
\end{eqnarray}
where $V_3=2\pi^2$ is the area of $S^3$. 
An interesting feature here is that the finite mass black hole solutions have two free parameters in the
event horizon initial data, which were taken to be $V(r_h)$ and $\tilde V(r_h)$.
As a result, and in contrast to the case of globally regular solutions, the two electric charges $q$ and $\tilde q$
are independent for black holes. 
This leads to a much richer parameter space of solutions.

Our numerical results provide evidence  
for the existence of finite mass
black hole solutions of the EYMCS system with a set of four notrivial gauge functions ($i.e.$ for the full group $SO(6)$).
The profile of a generic black hole corresponding to $\kappa=0.2$ and $r_h=0.5$ is presented in Figure 6 (left) for 
$\tilde q = 0.5,~q = -1.5$. 
%
%
\begin{figure}[ht]

\hbox to\linewidth{\hss%
	\resizebox{8cm}{6cm}{\includegraphics{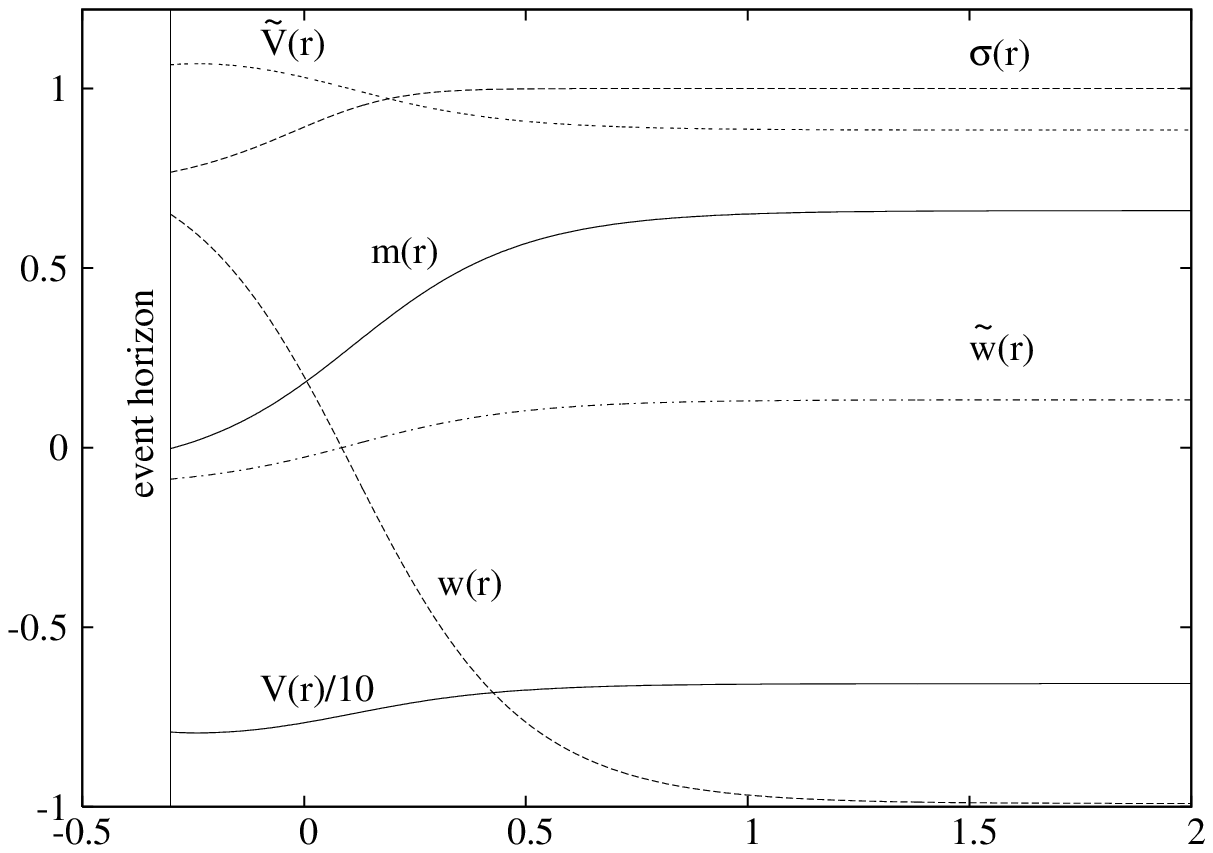}}
\hspace{5mm}%
        \resizebox{8cm}{6cm}{\includegraphics{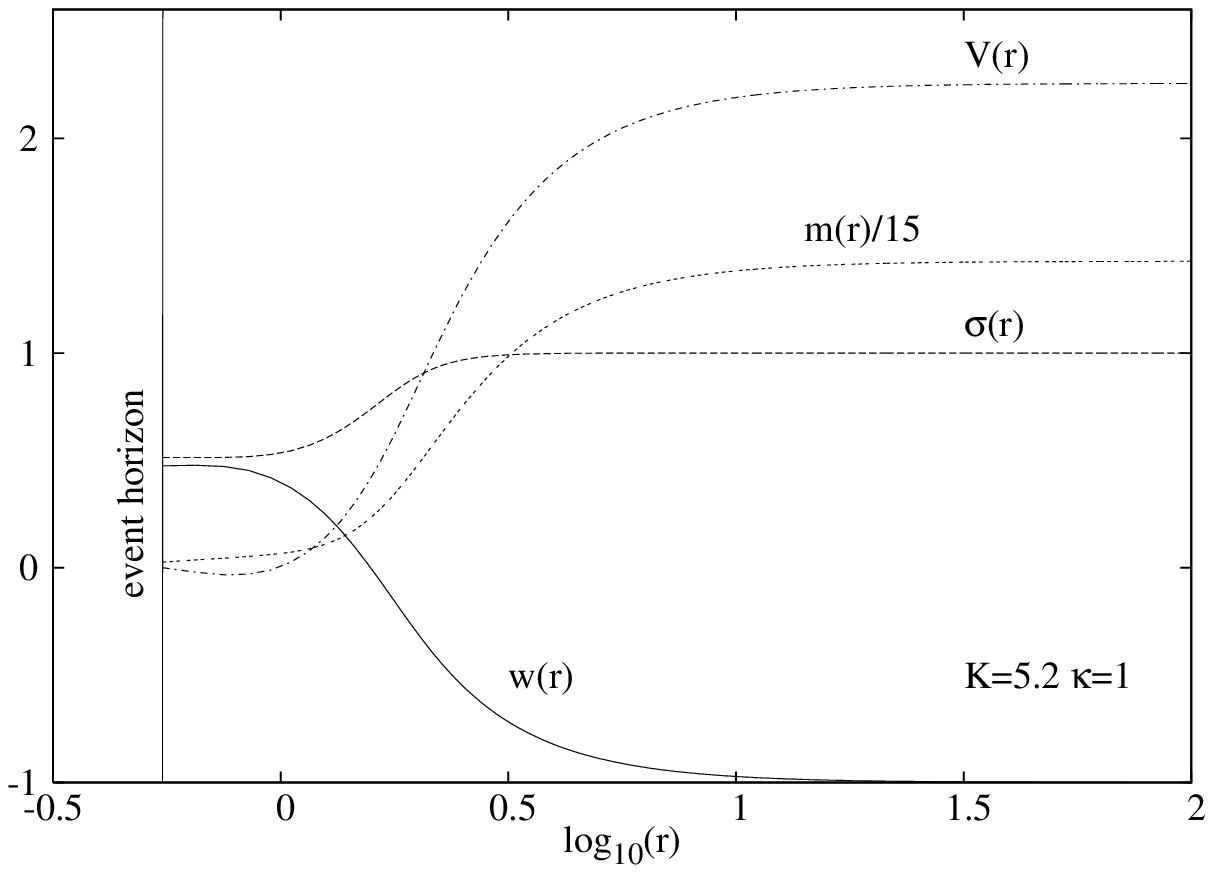}}	
\hss}

\caption{
{\small
{\it Left:} The profiles of a typical non extremal 
black hole solution of the $SO(6)$ model is presented as a function of the radial coordinate $r$.
{\it Right:} An extremal black hole solution for  our $particular\ truncation$ of the full model.
}
}
\label{fig46}
\end{figure}

\begin{figure}[ht]
\hbox to\linewidth{\hss%
	\resizebox{8cm}{6cm}{\includegraphics{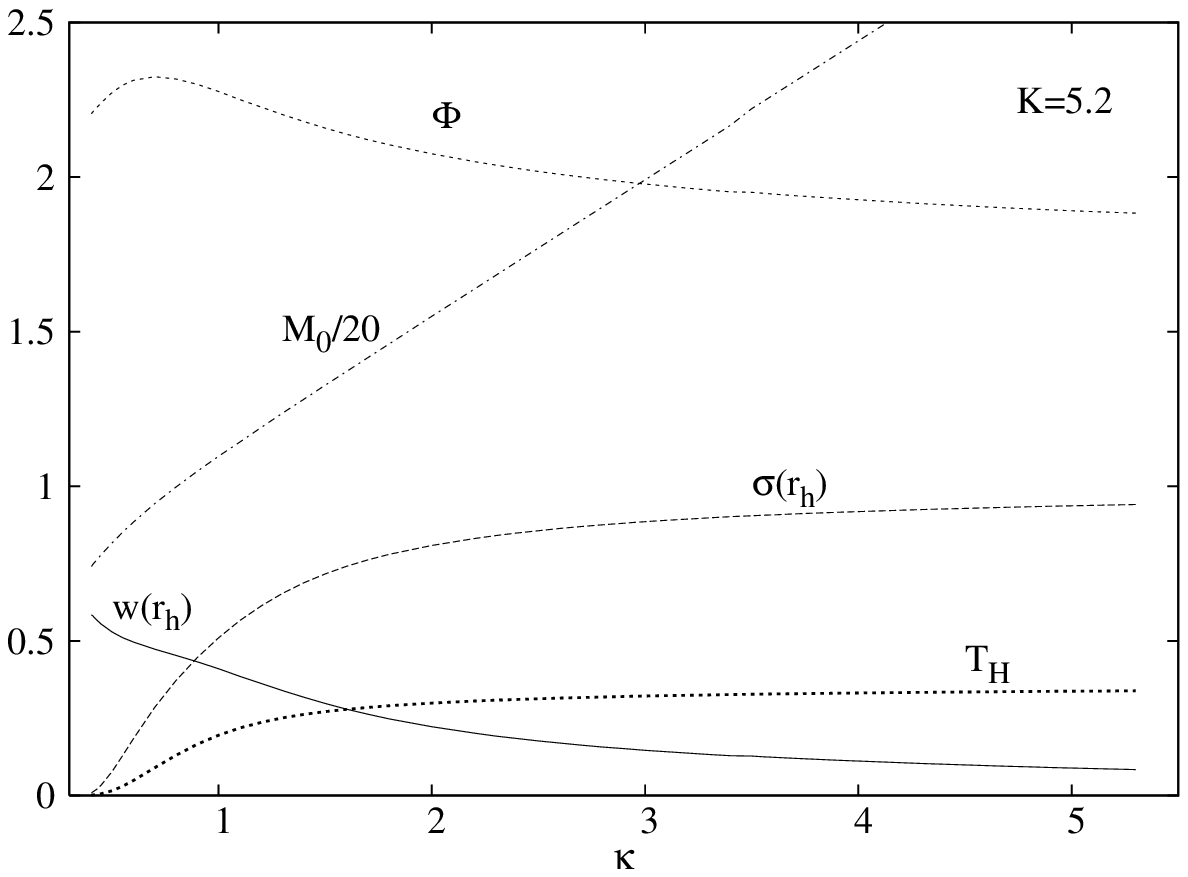}}
\hspace{5mm}%
        \resizebox{8cm}{6cm}{\includegraphics{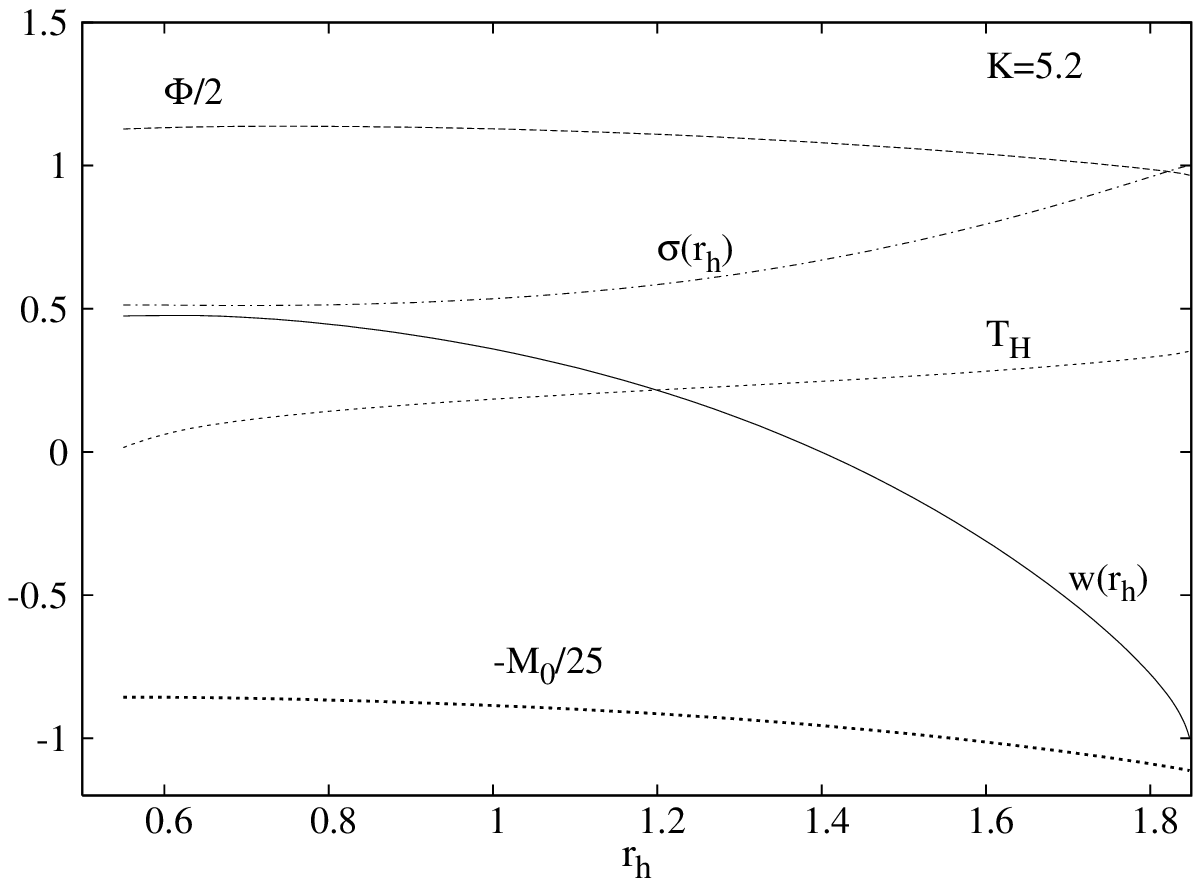}}	
\hss}

\caption{
{\small
The mass parameter $M_0$, the value at the horizon of the metric function $\sigma(r)$
and the magnetic gauge potential $w(r)$, the electrostatic potential $\Phi$ and the Hawking temperature
are plot as a functions of the CS coupling constant $\kappa$  (left) and of the event horizon radius $r_h$ (right). 
 }
 }
\label{fig47}
\end{figure}
\begin{figure}[ht]
\hbox to\linewidth{\hss%
	\resizebox{8cm}{6cm}{\includegraphics{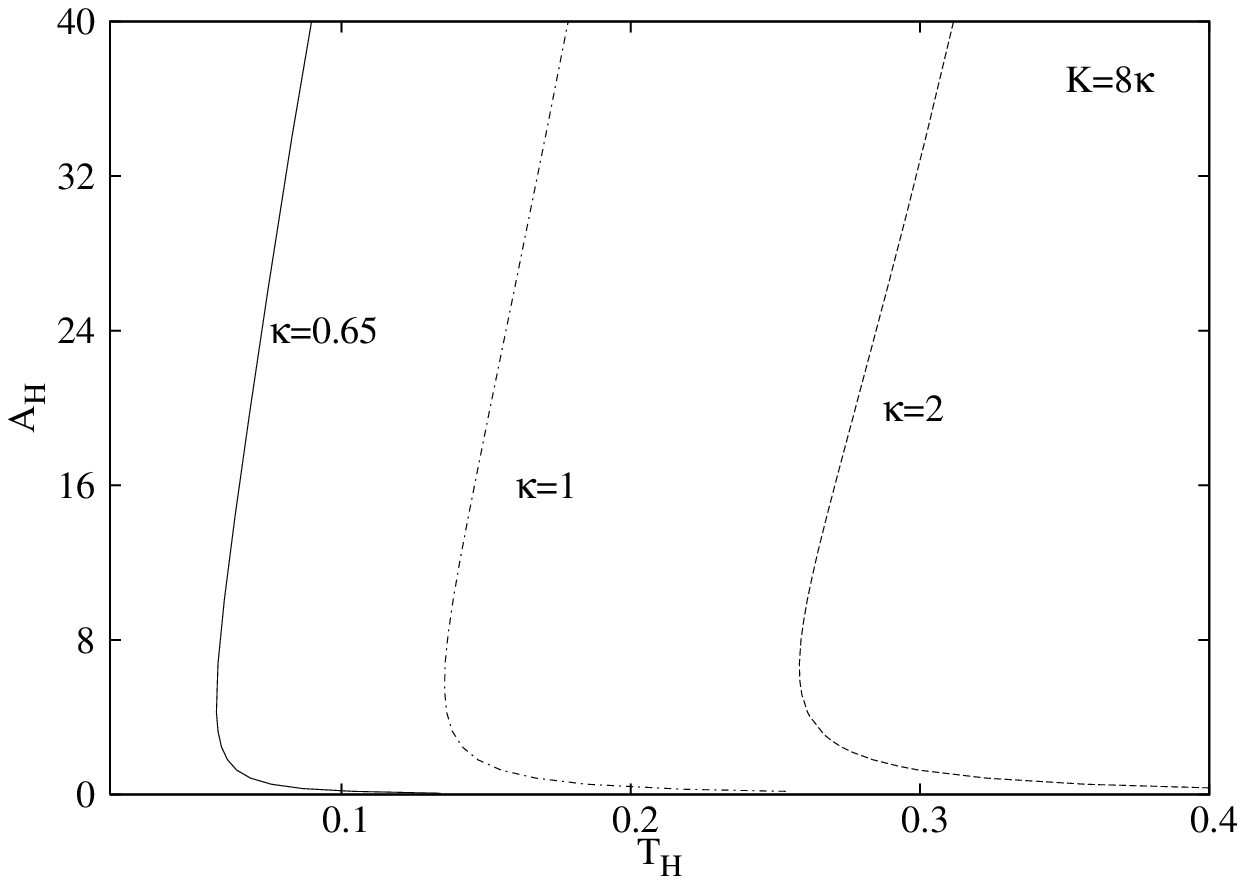}}
\hspace{5mm}%
        \resizebox{8cm}{6cm}{\includegraphics{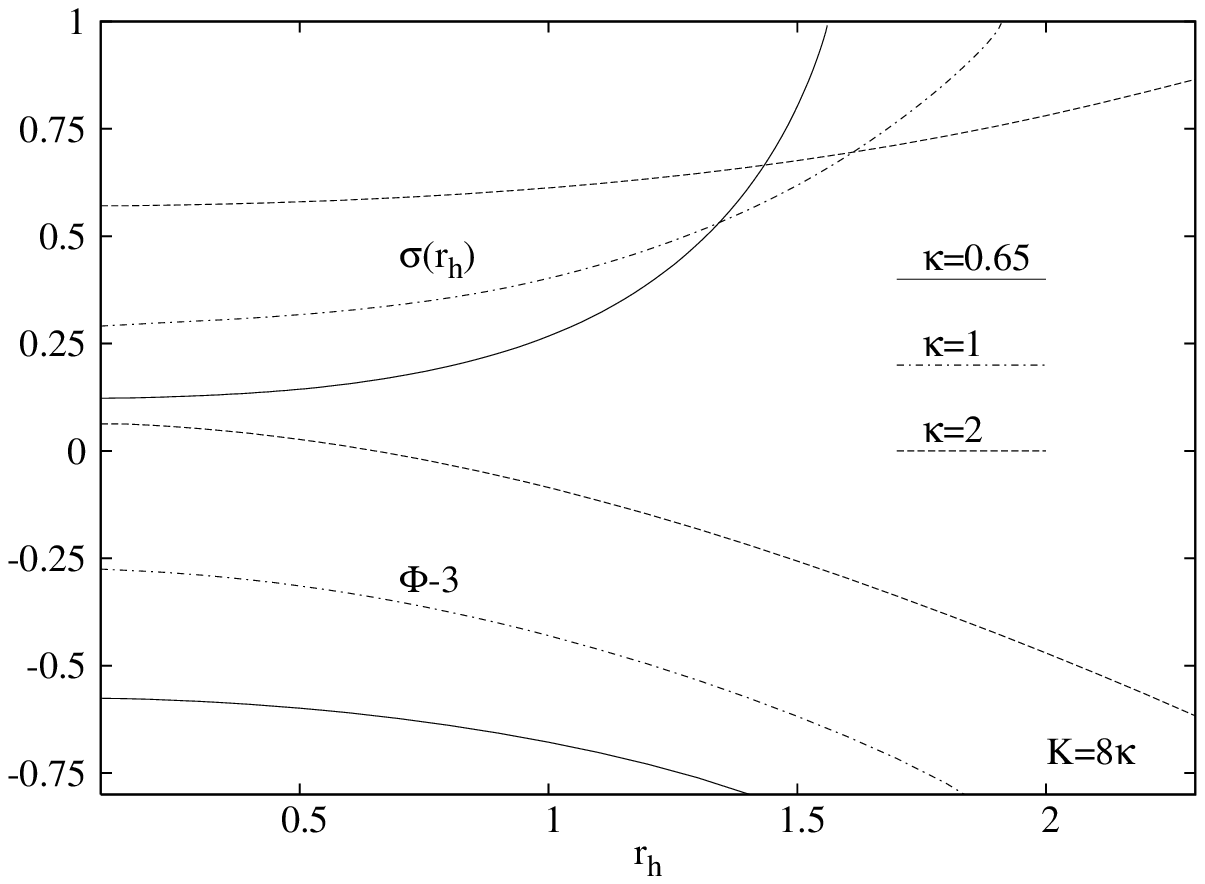}}	
\hss}
\caption{
{\small 
The temperature-entropy diagram (left) and the value at the horizon of the metric function $\sigma(r)$ and of the
electrostatic potential $\Phi$ (right) are ploted as functions of the event horizon radius $r_h$ for three different values of
the Chern-Simons coupling constant $\kappa$. These configurations have an integration constant $K=8 \kappa$, and approach the globally regular
particle-like solutions as $r_h\to 0$.
}
 }
\label{fig48}
\end{figure}

However, given the large number of free input parameters, 
we did not attempt a systematic study of these solutions, concentrating instead on the simpler case of 
the $SU(2)\times U(1)$ truncated model. An interesting feature here is that solutions with AdS asymptotics
exist only for the limited interval $ \kappa_{min} \leq \kappa \leq \kappa_{max}$ of $\kappa$.
The limits of this interval depend on the values of $r_h$ and $K$. Clearly, the range of the 
electric charge of these solutions is also bounded.
These features are illustrated in Figure 7 (left), for black hole solutions with a fixed horizon radius $r_h=1$.
One can see that as $\kappa \to \kappa_{min}$ 
the value at the horizon of metric function $\sigma(r)$ 
tends to zero, as does also the Hawking temperature, while other quantities stay finite.
The behaviour of solutions for large $\kappa$ is less clear, the accuracy decreasing with $\kappa$.
However, the numerical results seem to indicate that this branch ends in a critical solution with $\sigma(r_h)$
close to one and a finite nonzero value of $T_H$.
Unfortunately, the study of solutions for $ \kappa \to \kappa_{max}$ is a difficult task and  
the general picture may be much more complicated.
For example, we have noticed the existence there of a secondary branch of solutions, which 
are close to a finite mass extremal configuration with nontrivial gauge fields.
A systematic study of these aspects would require a different parametrisation of the metric
line element than (\ref{metric-gen}), and is beyond the scope of this work.

One can also keep $\kappa$ and $K$ fixed and vary the value of the event horizon radius. As noted above, 
choosing the values of $\kappa,K$ fixes also the electric charge, $i.e.,$ these black holes are in a canonical ensemble.
Our numerics indicate that for any $K$, the value of the gauge field potential at the horizon decreases with $r_h$.
For large enough values of the event horizon radius, the solutions become essentially RNAdS black holes,
with $w(r)$ being close to the value $-1$ everywhere, with the non-Abelian magnetic field vanishing.

However, the picture for small enough values of the horizon radius
depends crucially on the value of the integration constant $K$ in the $V$-equation (\ref{FI})
($i.e.$ on the Abelian electric charge).
Starting with the special value $K=8\kappa$, we plot in Figure 8 a number of relevant features of the solutions, 
for three different values of the CS coefficient $\kappa$.
One can see that these non-Abelian black holes behave in a similar way to the vacuum AdS solutions.
The small black holes are thermally unstable, the entropy being a decreasing function of the Hawking
temperature. They become thermally stable for large enough values of the event horizon radius.
In the limit $r_h\to 0$, these black holes 
approach the set of globally regular particle-like solutions with $\tilde V(r)=\tilde w(r)=0$, discussed above .

The picture is very different when choosing instead $K \neq 8\kappa$ (see Figure 7 (right)).
As an interesting new feature, here an extremal black hole solution is approached for a critical value of $r_h$.
In this case the horizon is degenerate ($i.e.$, $N(r)$ has a double root: $N(r_h) = N'(r_h) = 0$) and the near
horizon geometry is $AdS_2 \times S^3$. 

As $r\to r_h$ one finds the approximate form
of the solution in the near horizon region: 
\begin{eqnarray}
&&N(r)=N_2(r-r_h)^2+O(r-r_h)^3 ,~~\sigma(r)=\sigma_h-\frac{3\sigma_h w_1^2}{2r_h}(r-r_h)+O(r-r_h)^2,
\\
&&w(r)=w_h+w_1(r-r_h)+O(r-r_h)^2,~~~V(r)=V_h-\frac{\sigma_h w_h}{2\kappa r_h}(r-r_h)+O(r-r_h)^2,
\end{eqnarray}
where
\begin{eqnarray}
&&w_1= 
\frac{32 \kappa^2 r_h \ell^2w_h(1-w_h^2)}
{-3\alpha^2 r_h^2\ell^2 w_h^2+384 \kappa^4 \ell^2 (1-w_h^2)^2
-4\kappa^2(24r_h^4-4r_h^2\ell^2(w_h^2-2)+3\alpha^2\ell^2(1-w_h^2)^2)}
\\
\nonumber
&&N_2=\frac{24 r_h^4+8r_h^2\ell^2-3 \alpha^2\ell^2(1-w_h^2)^2)}{2r_h^4\ell^2}.
\end{eqnarray}
The parameters $r_h$ and $w_h$ in the above relation 
are solutions of the equations
\begin{eqnarray}
&&\frac{32r_h^4}{\ell^2}-12\alpha^2 (1-w_h^2)^2+r_h^2(16-\frac{\alpha^2 w_h^2}{\kappa^2})=0,
\\
\nonumber
&&2K \kappa +w_h(r_h^2+8\kappa^2(w_h^2-3))=0.
\end{eqnarray}
Recalling that $K=2(q-4\kappa)$, it follows that all event horizon boundary data (except $\sigma(r_h)$)
are fixed by the  $\kappa,~\ell$ and the electric charge $q$. (Note the analogy with the extremal Abelian solution case.)
This extremal solution differs from the RNAdS one, presenting 
non-Abelian magnetic hair and a nontrivial metric function $\sigma(r)$ (see Figure 6 (right)).
%
%

As expected, the near horizon structure of the extremal solutions can be extended to a full AdS$_2\times S^3$
solution of the field equations. This configuration has a line element
\begin{eqnarray}
\label{ex1}
ds^2=\frac{dr^2}{1-\frac{\Lambda_1 r^2}{6}}
+r_0^2 d\Omega_3^2-(1-\frac{\Lambda_1 r^2}{6})dt^2,
\end{eqnarray}
and the matter fields
\begin{eqnarray}
w=w_0,~~V(r)=V_0-\frac{w_0}{2\kappa r_0}(r-r_h).
\end{eqnarray}
For $w_0^2\neq 1$, this is a non-Abelian solution, with the gauge field living on the three-sphere. 
The parameters $\Lambda$, $w_0$ and the radius $r_0$ of the $S^3$ are constrained by the relation
\begin{eqnarray}
\Lambda=\frac{3}{r_0^2}-\frac{\pi G}{e^2}\frac{r_0^2 w_0^2+12 \kappa^2 (w_0^2-1)^2}{\kappa^2 r_0^4},
\end{eqnarray}
the value of the AdS$_2$ cosmological constant $\Lambda_1$ in (\ref{ex1}) being
\begin{eqnarray}
\Lambda_1=6
\left(
 \Lambda-\frac{1}{r_0^2}-\frac{\pi G}{e^2}\frac{r_0^2 w_0^2+4 \kappa^2 (1-w_0^2)^2}{\kappa^2 r_0^4} 
\right).
\end{eqnarray}

\section{Global charges}

\subsection{The mass and boundary stress tensor}

The action and mass of these AdS$_5$ non-Abelian configurations is computed by using a boundary
counterterm prescription. As found in \cite{Balasubramanian:1999re},
the following counterterms are sufficient to cancel divergences in five dimensions, 
for Schwarzschild-AdS black hole solution (in this Section we restore
the $8\pi G$ and $e$ factors in the expressions):
\begin{eqnarray}
\label{ct}
I_{\rm ct}=-  \frac{1}{8\pi G}\int_{\partial {\cal M}}d^{4}x\sqrt{-h}\Biggl[
\frac{3}{ \ell}+\frac{ \ell}{4}\rm{R}
\Bigg]\ ,
\end{eqnarray}
with $\rm{R}$ the Ricci scalar for the boundary metric $h$.

(Note also that, as usual, to ensure well-defined Euler-Lagrange field equations,
one adds to the action
(\ref{action}), the
Gibbons-Hawking surface term \cite{Gibbons:1976ue}~~ $I_{{\rm
surf}}=-\frac{1}{8\pi G}\int_{\partial\mathcal{M}} d^{4}x\sqrt{-h}K$,
where $K$ is the trace of the extrinsic curvature for the boundary
$\partial\mathcal{M}$.)
However,  in the presence of matter fields, additional counterterms may be needed to regulate the action
\cite{Taylor-Robinson:2000xw}, which is also the case  
for the generic non-Abelian solutions discussed in the previous Sections\footnote{
The geometric counterterm (\ref{ct}) regularises also the 
action of the RNAdS$_5$ black hole solution. However,
this does not hold for any $d=5$ solutions of the Einstein-Maxwell-$\Lambda$ system.
An interesting example here are the AdS black strings 
with a magnetic $U(1)$ field \cite{Bernamonti:2007bu}, in which case one has to consider an additional
matter counterterm on the form (\ref{Ict-mat}). }.
 
This divergence is cancelled by a supplementary  counterterm of the form (with $a,b$ boundary indices):
\begin{eqnarray}
\label{Ict-mat}
I_{ct}^{YM}=
- 
\log(\frac{r}{\ell})
 \int_{\partial {\cal M} }d^{4}x\sqrt{-h }\frac{\ell}{2e^2 }
 ~{\rm Tr}\{ F_{ab}F^{ab}\}~.
\end{eqnarray}
Note that this term is identically zero for the solutions 
with $w^2+\tilde w^2 \to 1$, $\tilde V w - V \tilde w \to 0$.

Using these counterterms  and the Gibbons-Hawking boundary term, 
one can construct a divergence-free boundary stress tensor ${\rm T}_{ab}$ 
\begin{eqnarray}
\label{TAB-mat}
 {\rm T}_{ab}= 
 \frac{1}{8\pi G}(K_{ab}-Kh_{ab}-\frac{3}{\ell}h_{ab}+\frac{\ell}{2} E_{ab})
-\frac{2\ell}{e^2} \log(\frac{r}{\ell})
~{\rm Tr}\{F_{ac}F_{bd}h^{cd}-\frac{1}{4}h_{ab}F_{cd}F^{cd}\}~,~~{~} 
\end{eqnarray}
where $E_{ab}$ and $K$ are the Einstein tensor and the trace of the extrinsic curvature $K_{ab}$ for the induced
metric of the boundary, respectively. In this approach, the mass ${\rm M}$ 
of the solutions is the conserved charge associated with the Killing vector 
$\partial /\partial t$  \cite{Balasubramanian:1999re}.
A straightforward computation leads to the following simple result for the mass of the generic EYMCS solutions:
\begin{eqnarray}
\label{mass}
{\rm M}=\frac{3V_3 M_0}{16\pi G}+M_c, ~~{\rm with ~~}M_c =
\frac{3 V_3 \ell^2}{64\pi G}~.
\end{eqnarray}
For the case of black hole solutions of the truncated $SU(2)\times U(1)$ model,
we have found that ${\rm M}$
coincides within the numerical accuracy 
with the  mass computed from the first law of thermodynamics, up to the constant
term $M_c$ which is usually interpreted as the mass of the pure global AdS$_5$.

From the AdS/CFT correspondence, we expect 
the non-Abelian hairy black holes to be described by
some thermal states
in a dual theory  
 formulated in a metric background given by 
 \begin{eqnarray}
 \gamma_{ab}dx^a dx^b=-dt ^2+\ell^2 (d\psi^2+\sin^2\psi (d\theta^2+\sin^2\theta d\varphi^2)), 
\end{eqnarray}
where $\psi,\theta,\varphi$ are the usual polar angles parametrizing $S^3$.

The matter fields in the dual CFT would interact with a background
non-Abelian field, whose expression, as read from  (\ref{a02}), (\ref{asimpt-div}) is
\begin{eqnarray}
\nonumber
A_t&=& \frac{1}{e}\big( 
\tilde V_0\left[\sin\psi\,\sin\ta\left(\Si_{16}\cos\vf+\Si_{26}\sin\vf\right)+\sin\psi\cos\ta\,\Si_{36}
+\cos\psi\Si_{46}\right]-V_0\,\Si_{56}
 \big),
\label{a0}
\\
\nonumber
A_{\psi}&=&\frac{1}{e}\big(
(1+w_0)\left[\sin\ta(\Si_{14}\cos\vf+\Si_{24}\sin\vf)+\cos\ta\,\Si_{34}\right]\nonumber\\
&~~&+\tilde w_0\left[\cos\psi\sin\ta(\Si_{15}\cos\vf+\Si_{25}\sin\vf)+\cos\psi\cos\ta\Si_{35}-\sin\psi\Si_{45}\right]
 \big),
\label{apsi}
\\
\nonumber
A_{\ta}&=&\frac{1}{e}\big(
(1+w_0)\sin\psi\big[\sin\psi(\Si_{13}\cos\vf+\Si_{23}\sin\vf)-\sin\psi\cos\ta\,\Si_{14}\cos\vf
\\
\nonumber
&& +\cos\psi\cos\ta\,\Si_{24}\sin\vf+\cos\psi\sin\ta\,\Si_{34}\big]
 +\tilde w_0\sin\psi\left[\cos\ta(\Si_{15}\cos\vf+\Si_{25}\sin\vf)-\sin\ta\,\Si_{35}\right]
  \big),
\label{ata}
\\
A_{\vf}&=&
\frac{1}{e}\big(
-(1+w_0)\sin\psi\big[\sin\psi\sin^2\ta\,\Si_{12}+\sin\psi\sin\ta\cos\ta(\Si_{13}\sin\vf-\Si_{23}\cos\vf)
\nonumber
\\
&& +\cos\psi\sin\ta(\Si_{14}\sin\vf-\Si_{24}\cos\vf)\big]
 -\tilde w_0\sin\psi\sin\ta(\Si_{15}\sin\vf-\Si_{25}\cos\vf)
 \big),
\nonumber
\label{avf}
\end{eqnarray}
We note that this is still fully an $SO(6)$ gauge field.

The expectation value $<\tau^{a}_b>$ of the dual CFT stress tensor
can be calculated using the  relation \cite{Myers:1999qn}
\begin{eqnarray}
\label{r1}
\sqrt{-\gamma}\gamma^{ab}<\tau_{bc}>=
\lim_{r \rightarrow \infty} \sqrt{-h} h^{ab}{  {\rm T}}_{bc}.
\end{eqnarray}
Employing also (\ref{TAB-mat}), we find the finite and covariantly 
conserved stress tensor
(with $x^1=\psi,~x^2=\theta,~ x^3=\varphi,~x^4=t$)
\begin{eqnarray}
\label{st1}
8 \pi G <\tau^{a}_b> = 
\frac{1}{2\ell}
\big( 
 \frac{M_0}{\ell^2} +\frac{1}{4}
\big)\left( \begin{array}{cccc}
1&0&0&0
\\
0&1&0&0
\\
0&0&1&0
\\
0&0&0&-3
\end{array}
\right)
-\frac{4\pi G( (w_0^2+\tilde w_0^2-1)^2+\ell^2(\tilde V_0 w_0-V_0 \tilde  w_0)^2 )}{e^2\ell^3}
\left( \begin{array}{cccc}
1&0&0&0
\\
0&1&0&0
\\
0&0&1&0
\\
0&0&0&0
\end{array}
\right)
.
\end{eqnarray}
Different $e.g.$ from the case of Reissner-Nordstr\"om-AdS Abelian solutions, this stress tensor has a nonvanishing trace.
Moreover, for the physically relevant case of solutions with a finite electric charge ($i.e.$ $|\vec \phi \times \vec \chi|\to 0$
asymptotically) one finds that $ <\tau^{a}_a>={\cal A}_{YM}=-3 (w_0^2+\tilde w_0^2-1)^2 /(2\ell^2 e^2)$.
This agrees with the general results \cite{Coleman:1970je}, \cite{Blau:1999vz},
\cite{Taylor-Robinson:2000xw} on the trace anomaly in the presence of an external gauge field, 
${\cal A}_{YM}={\cal R} F_{(0)}^2$, the coefficient ${\cal R}$ being related to the charges of the fundamental
constituent fields in the dual CFT.

\subsection{Electric charge(s)}
For the solutions with  $|\vec \phi \times \vec \chi | \to 0$ ($i.e.$ $(\tilde V_0 w_0-V_0 \tilde  w_0)|_{r\to \infty}=0$)
the coefficients of the $1/r^2$ terms in the asymptotic expansion of the electric potentials are finite.
Thus, from the Gauss flux theorem one can   formally  define  the electric charge 
\begin{eqnarray}
\label{r2}
 Q_E=
\oint_{\infty} dS_k\sqrt{-g} F^{k t}= \frac{4\pi^2}{e}(q \Sigma_{56}+\tilde q \Sigma_{12})\,,
\end{eqnarray}
 which is clealy not gauge invariant. (This is a generic 
problem for the definition of the non-Abelian charges in the absence of a
Higgs field, see $e.g.$ \cite{Volkov:1999cc}). 

Perhaps a more proper definition can be given following the reasoning in 
\cite{Sudarsky:1992ty}. In this approach one starts by evaluating the quantity $\mbox{Tr} \{ F_{i t}F^{i t}\}$,
 \[
\sqrt{-g} \mbox{Tr}\,\{ F_{i t}F^{i t} \}= \sqrt{-g} \mbox{Tr}\, \{D_iA_t\,F^{it}\}
=\pa_i (\sqrt{-g} \mbox{Tr}\,\{ A_t\,F^{it}\}) -\sqrt{-g} \mbox{Tr}\,\{ A_t\,D_iF^{it} \}\,.
\]
 Using the Gauss' law equation, we find that the contribution of the electric field to the total mass then is
\be
\label{elecen}
E_e=-\frac{1}{e^2}\int dS_k \sqrt{-g}\,\vec\chi\cdot D_r\vec\chi +4V_3 \frac{\ka}{e^2} \,\int dr\,(|\vec\f|^2-1)\,\vec\chi\cdot D_r\vec\f~g^{tt}
\ee
Subject to the truncations this expression simplifies; the integral in the second term must be evaluated using
the numerical solution, while the surface integral in the first term can be evaluate using only the asymptotic
values of the functions.

In the absence of the Chern-Simons term, $i.e.$, when $\ka=0$,
 the contribution of the electric field to the total mass can be written as 
\[
 E_e=-\oint_{\infty} dS_k \sqrt{-g} \mbox{Tr}\,\{ A_t\,F^{kt}\}\,=  \bf{ Q_E \Phi}~,
\]
where
\begin{eqnarray}
{ \bf \Phi}= \sqrt{\mbox{Tr}\,\{A_t A_t \}}=
 \sqrt{V_0^2+\tilde V_0^2}, ~~{\bf Q_E}= \frac{4\pi^2}{e} \frac{V_0 q+\tilde V_0 \tilde q}{\sqrt{V^2+\tilde V^2}},
\end{eqnarray}
 are the electrostatic potential and the electric charge, respectively.
However, one can extend this definition of ${\bf \Phi}$ and ${\bf Q_E}$ to solutions of the EYMCS system. 
This applies to both $generic$ and $special$ configurations (note that
${\bf \Phi}=\Phi$, ${\bf Q_E}=4\pi^2 (\sin \alpha~q+\cos \alpha~\tilde q)/e$ for finite mass solutions). 
  
For finite mass solutions, following \cite{Bartnik:1988am}, one can also define an effective non-Abelian charge
$Q_{eff}$ by the asymptotic behaviour
of the metric function $N(r)$, which, to order $1/r^4$ is similar to that of the RNAdS solution:
\begin{eqnarray}
\label{Qeff}
N(r)=1+\frac{r^2}{\ell^2}-\frac{M_0}{r^2}+\frac{Q_{eff}^2}{r^4}+\dots,
\end{eqnarray}
$i.e.$
\begin{eqnarray}
\label{Qeff2}
 Q_{eff}= \frac{\alpha\sqrt{(q^2+\tilde q^2)\ell^2+3(w_2^2+\tilde w_2^2))}}{\ell}.
\end{eqnarray}

 Concerning a definition of a "magnetic'' flux, the only natural quantity we have at our disposal for this purpose
is the Chern--Pontryagin density, which we know is the leading and sole contributing term to the topological charge
("magnetic'' flux) of the monopole in $4+1$ dimensions~\cite{O'Brien:1988xr}. There however the gauge group is $SO(4)$
and the model features a iso-four-vector Higgs field.

This quantity can be calculated easily for the $SO(6)$ Ansatz employed in this paper
\be
\label{mag}
\vep_{ijkl}\,\mbox{Tr}\,
\big \{
F^{ij}\,F^{kl}
\big \}
=-\frac{4!}{r^3}\,(|\vec\f|^2-1)\,\mbox{Tr}\left(D_r\,\f^3\,\Si_{56}+(D_r\,\f\vep)^M\,\Si_{M4}\right)\,,
\ee
 which vanishes, $i.e.$, the candidate for a magnetic charge for the solutions found in the present work
equals zero identically. 

\section{ Further remarks }
On general grounds, one expects that extending the known classes of solutions of the $d=5$ supergravity
to a non-Abelian gauge group would lead to a variety of new physical effects. 

This work has been aimed as a first step towards constructing 
the non-Abelian solutions of the maximal $d=5$ gauged supergravity.
Restricting to the simplest case of static, spherically symmetric solutions,
we have proposed a suitable Ansatz for the gauge fields and presented numerical evidence for the existence of both 
particle-like and black hole solutions. Our systematic description of the black holes is restricted to a certain
truncation of the full $SO(6)$ model, with the sole purpose of rendering the numerics practicable. In this limited
context, we have also found extremal black holes. 

As a consequence of the presence of a negative cosmological constant in the model, we have recovered the qualitative
properties of the solutions to the usual EYM-$\Lambda$ model in $3+1$ dimensions
\cite{Winstanley:1998sn}, \cite{Bjoraker:2000qd}. Notably, some of our solutions to which we have referred as $generic$,
are characterised by arbitrary asymptotic values of the potentials parametrising the gauge field. 
Also, these solutions share another property
with those of the $3+1$ dimensional EYM model, namely that the shooting parameters 
involved take on a continuum of values. Unlike the latter
however, their masses turn out to be divergent in our $4+1$ dimensional case. 
This is expected on the basis of the Derrick-type scaling argument.
However,  we have proposed a regularisation procedure for the  mass of these solutions, in the context of the AdS/CFT
correspondence. As far as these $generic$ solutions are concerned, the presence of the Chern--Simons term makes no
qualitative difference. 

Perhaps  the most interesting feature of the EYMCS model is the
existence of finite mass solutions.  We have referred to these as $special$ solutions and they contrast with the
$generic$ ones in that the shooting parameters involved take on a discrete set of values. The $special$ solutions exist
only when the non-Abelian Chern--Simons term is present.  

Concerning the physical context of our results, it is in order to make
several remarks on the issue of the more general solutions of the 
$d=5, ~{\cal N}=8$ gauged supergravity. This model contains in addition twenty scalars, which are represented
by a symmetric unimodular tensor. 
These scalars have a nontrivial potential approaching a constant negative value at infinity which
fixes the value of the effective cosmological constant. No obvious consistent truncation of this sector 
seems to exist for a gauge
group $SO(6)$ and strictly speaking one should work with the full set of scalars.
In principle, at least when this general model is subjected to spherical symmetry, the one dimensional subsystem
resulting from the aplication of the Ansatz here can be studied using the same methods as in this paper, 
$i.e.$, the solutions can be constructed by solving a boundary value problem.
One can in that case find the approximate expressions at the origin or event horizon and at infinity.
The only obstacle to this task that we see at this moment is the huge complexity of the ensuing equations.
Based on the results in this paper, we expect the parameter space of the full $SO(6)$ solutions of the $d=5,
~{\cal N}=8$ model to be very rich. Inclusion of the scalar sector will lead to many new free parameters in the
asymptotics and will make any attempt to classify the solutions difficult 
(involving   numerous  different ways of approaching a constant negative value at infinity
for the scalar potential).

The $generic$ non-Abelian solutions on the other hand will always present a nonvanishing
magnetic gauge field on the boundary which appears
as a background for the dual theory. (This feature is independent of the presence or absence of scalars.)
Thus the expectation value of the dual CFT stress tensor will contain a part which is similar to (\ref{st1}).
Configurations with vanishing non-Abelian magnetic field on the boundary, and finite mass, should exist as well,
being supported by the CS term.
Also, similar to the case of four dimensional EYM  non-Abelian solutions in \cite{Mann:2006jc}, 
the existence of both spherically symmetric globally regular
and hairy black hole solutions with the same set of data at infinity raises the question as to how 
the dual CFT is able to distinguish between these different bulk configurations. 

Concerning future developments of this work,
we expect a much richer structure of the non-Abelian solutions 
to be found when relaxing the spacetime symmetries.
For example, one may envisage the existence of asymptotically AdS$_5$ solutions which are static
and non-spherically symmetric, generalising the configurations in \cite{Radu:2007jb}.
Particularly interesting would be to approach the issue of EYMCS rotating solutions.
To our knowledge, the only $d>4$ rotating solutions with non-Abelian fields known so far in the literature
are the $d=5$ EYM-$SU(2)$ black holes in \cite{Brihaye:2007tw}.
However, the mass of these solutions diverges logarithmically.
It is likely that the inclusion of a CS term will lead to finite mass solutions also in that case.

Another very natural direction to be explored is the case of zero cosmological constant,
which would be outside the context of ${\cal N}=8$ supergravity but would
nonetheless be technically very interesting. For example, this would afford a comparison with the corresponding 
$3+1$ dimensional asymptotically flat EYM solutions in \cite{Bartnik:1988am},
which involve only a discrete spectrum of the shooting parameters.
\\
\\
{\bf\large Acknowledgements} \\
One of us (D.H.Tch.) would like to thank Werner Nahm and Ruben Manvelyan for
useful discussions. E.R. and D.H.Tch. are pleased to acknowledge enlightening
and encouraging discussions with Michael Volkov.
This work is carried out in the framework of Science Foundation Ireland (SFI) project
RFP07-330PHY. 
YB is grateful to the
Belgian FNRS for financial support.
 The work of ER was supported by a fellowship from the Alexander von Humboldt Foundation.

\begin{small}

\end{small}

\end{document}